\documentclass[11pt,a4paper]{article}
\usepackage{jheppub}

\usepackage{physics}
\usepackage{tensor}
\usepackage{amsmath}
\usepackage{amssymb}
\usepackage[colorlinks=true]{hyperref}
\usepackage{xcolor}
\usepackage{cancel}
\usepackage{ytableau}

\usepackage{titlesec}
\makeatletter
\g@addto@macro\bfseries{\boldmath}
\makeatother

\ytableausetup{boxsize=0.3em,centertableaux}

\usepackage{framed}
\usepackage[most]{tcolorbox}
\usepackage{xcolor}
\colorlet{shadecolor}{gray!10}

\preprint{{Imperial-TP-2025-CH-8}\\ \rightline{{UUITP-23/25}}}

\title{Gravitational currents and charges from conformal Killing-Yano tensors}

\author[a]{Chris Hull,}
\author[b]{Ulf Lindstr\"{o}m,}
\author[a]{Maxwell L. Vel\'{a}squez Cotini Hutt}

\affiliation[a]{The Blackett Laboratory, Imperial College London, Prince Consort Road, London, SW7 2AZ, UK}
\affiliation[b]{Department of Physics and Astronomy, Uppsala University, Box 516, SE-75120 Uppsala, Sweden and Centre for Geometry and Physics, Uppsala University, Box 480, SE-75106 Uppsala, Sweden}

\emailAdd{c.hull@imperial.ac.uk, m.hutt22@imperial.ac.uk, ulf.lindstrom@physics.uu.se}

\abstract{We construct a set of higher-form conserved currents on spacetimes admitting conformal Killing-Yano tensors. We find relations between these currents that allow the charge given by integrating one of these currents over a region to be re-expressed as an integral of a covariant quantity over the boundary. In many cases, only a non-covariant form of the boundary integral was previously known. For a special class of these currents the conservation does not rely on field equations, so they give conserved topological charges in any gravitational theory. We discuss the relation of our currents to the Komar current and derive several new properties of conformal Killing-Yano tensors. We study a number of applications of our construction to charges of black hole solutions of Einstein-Maxwell theory, and D-brane solutions of type II supergravity.
}

\begin{document}
\pagestyle{myplain}
\maketitle

\section{Introduction}

The construction of conserved charges in  generally covariant theories requires spacetimes with extra structure.
Spacetimes with an asymptotic region with asymptotic Killing vectors allow the construction of \emph{asymptotic} charges. 
The fact that there are only asymptotic global charges in general is a reflection of the fact that gravitational energy, momentum, angular momentum, etc.~are non-local in generally covariant theories.
On asymptotically flat spacetimes, for example, constructions of this type lead to the ADM \cite{Arnowitt:1962hi} and BMS \cite{Bondi:1962px, Sachs:1962wk} charges given by integrals  at spatial and null infinity respectively.
Spacetimes with an isometry, on the other hand, have a \emph{local} conserved current that can be integrated over any codimension-1 surface to define a charge. 
For each isometry there is a globally defined Killing vector $k$ and the corresponding conserved current is given by contracting with the energy-momentum tensor, $j_\mu=T_{\mu\nu}k^\nu$. 
There are, however, spacetimes for which isometries alone do not yield all the local conserved currents and many works have discussed other structures which can be used to define conserved quantities in gravitational theories. A famous example is the Carter constant for motion around a black hole \cite{Carter:1968rr}. Such structures are not generic but are found in many spacetimes of interest, and our aim here is to explore a particular set of conserved currents and the charges defined by integrating them.

Commonly, the extra structures used to find conserved currents are  generalisations of Killing vectors. Examples include symmetric Killing tensors and anti-symmetric Killing-Yano (KY) tensors \cite{Yano1952, Kashiwada:1968fva}. These give rise to hidden symmetries of black hole solutions \cite{Kastor2004, Lindstrom:2022iec, Frolov:2008jr, Frolov:2017kze} and to the separability of PDEs on particular backgrounds \cite{Carter:1968ks, Carter:1968rr, Woodhouse:1975, Walker:1970un, Penrose:1973um, Carter:1977pq, Lunin:2020mxj, Krtous:2008tb}. For a review of the applications of these tensors, see \cite{Lindstrom:2022nrm}. A more general family of tensors, known as \emph{conformal Killing-Yano} (CKY) forms \cite{Tachibana1969OnSpace}, have been particularly relevant in constructing conserved quantities \cite{Ozkarsligil:2023avt, Benn:1996su, Lindstrom:2021qrk, Lindstrom:2021dpm, Jezierski:2020euo, Jezierski:2019xxl, Czajka:2017kzb, Jezierski:2014gka, Jezierski:2008zz, Jezierski:2007ym, Jezierski:2006fw, Jezierski:2005cg, Jezierski:2002mn, Jezierski:1994xm, Jezierski:1994cj, Kinoshita:2024wyr}. These tensors play a central role in the present work and are described in detail in section~\ref{sec:CKY}.

A useful playground to discuss conserved quantities is the Einstein theory linearised around a Minkowski background. This is a field theory on a fixed background and the problems regarding local definitions of energy are absent. The (generalised) symmetries of this theory have been discussed in many works \cite{Penrose:1982wp,Goldberg:1990ju,Jezierski:1994cj, Jezierski:2002mn,Aksteiner:2013rq,Hinterbichler2023GravitySymmetries, Benedetti2022GeneralizedGraviton, Benedetti:2023ipt, BenedettiNoether, Gomez-Fayren:2023qly, Hull:2023iny, Hull:2024mfb, Hull:2024xgo, Hull:2024ism, Hull:2024bcl, Hull:2024qpy}. In many of these, a central role is played by a set of conserved 2-form currents built from contractions of the linearised Riemann tensor with a CKY 2-form of the background Minkowski space. These currents yield charges defined on codimension-2 surfaces, generating 1-form symmetries. They have been related to a linearised version of the ADM charges \cite{Hull:2024xgo} and a set of magnetic gravitational charges \cite{Hull:2023iny, Hull:2024mfb} including NUT charges.

These currents have a natural generalisation to the generally covariant theory in spacetimes that admit a CKY tensor. As we will show, fully anti-symmetric conserved currents are constructed by suitable contractions of the CKY forms with the full Riemann tensor. Currents of this form are the focus of the present work. Given the role of the analogous charges in the linearised theory, one may expect that they are the correct objects to measure both `electric' charges (e.g.~masses and angular momenta) and `magnetic' charges (e.g.~NUT charge) in the full theory. We will confirm that this is the case.\footnote{Another generalisation of the currents in the linearised theory is to consider asymptotic symmetries of the generally covariant theory of the same form, with some particular boundary condition. We will return to this  in a future work.}

Given a CKY $p$-form, we find $(p-1)$-form conserved currents, as well as a $p$-form whose divergence yields the $(p-1)$-form current. Using terminology introduced in \cite{Hull:2023iny}, the conserved $(p-1)$-form currents are called \emph{primary currents} while the related $p$-forms are called \emph{secondary currents}.\footnote{The secondary current is also sometimes called a `potential' for the primary current.} Secondary currents are not uniquely defined as one can always add a co-closed term to them without changing their properties. However, the particular definition of the secondary current that we construct is naturally \emph{covariant}. The importance of using covariant quantities in topologically non-trivial situations has been highlighted in \cite{Marolf:2000cb, Hull:2024mfb}. Interestingly, we find that the conservation of the primary currents relies  on highly non-trivial integrability conditions of the CKY forms. We also take an alternative approach: given the existence of a secondary current, the associated primary current is conserved identically and we can use this to derive new integrability conditions on spacetimes admitting CKY forms.

The importance of the secondary currents is that they allow the charge defined by integrating a primary current over a suitable submanifold 
to be re-expressed as the integral of the secondary current over the boundary of that submanifold.
 The surface can be arbitrarily deformed and the value of the integral only changes when the surface passes over a charged object. In this sense the charge is a topological quantity.
This notion of a charge is similar to the one which has become standard in the discussion of generalised symmetries in QFT, following \cite{Gaiotto:2014kfa}, although we note that our discussion here is entirely classical. 
As the currents are covariant, this construction is coordinate independent, and we use the robust concept of a topological charge instead of other  notions of what it means for a charge to be conserved, which may have some coordinate dependence, e.g.\ they may involve  the existence of a global time coordinate.
As an example, we will discuss spacetimes admitting a rank-2 CKY tensor whose divergence is a timelike Killing vector. The associated charge is then a measure of the mass contained in a given region, and its value can jump whenever the integration surface is deformed through a region with 
$G_{\mu\nu}\neq 0$, which on-shell means a region with $T_{\mu\nu}\neq 0$.

We note that the functional form of the currents that we discuss is the same as those considered in \cite{Kastor2004}. However, in that work only KY forms were considered. We will see that the generalisation to include CKY forms gives a range of new applications of these currents, and is crucial for the construction of a covariant secondary current. 
As mentioned above, these currents in the generally covariant theory can be seen as a natural generalisation of currents in the theory linearised around Minkowski space. This, retrospectively,  provides a motivation for the form of the currents studied in \cite{Kastor2004}. 
In \cite{Kastor2004}, a charge was defined by integrating the current for a KY tensor over  a submanifold of a spatial hypersurface of an asymptotically flat spacetime, and then re-expressing the charge as a surface integral over a subspace of the sphere at spatial infinity using  the methods of \cite{Abbott:1981ff}. Here we instead integrate the current over a closed surface to define a topological charge.

Currents similar to the ones discussed here but which are only conserved asymptotically have also been studied via an adaptation of the procedure of \cite{Abbott:1981ff} on asymptotically flat and asymptotically anti-de Sitter backgrounds \cite{Cebeci2006}, while some subtler properties and generalisations have been studied recently in \cite{Lindstrom:2021qrk, Ozkarsligil:2023avt}. 
Some of these works have also discussed secondary currents but, to the authors' knowledge, the  present work gives the first  construction of covariant secondary currents.

As the charges associated with the currents are defined only on a submanifold, the tensors required to define the charge do not necessarily need to be globally defined. Instead, it would be sufficient for the relevant tensors to exist only on the submanifold where the charge is defined, or perhaps in a neighbourhood thereof. An example of such a situation is the twistorial construction of \cite{Penrose:1982wp}, which allows the construction of charges for a restricted set of submanifolds. While this significantly weakens the constraints imposed on the background spacetime for the CKY forms to exist globally, in this situation one cannot deform the surface over which the charge is defined arbitrarily. In the more modern language of \cite{Gaiotto:2014kfa}, a global symmetry should be associated with a surface which can be smoothly deformed in an arbitrary manner without affecting the charge if no sources are crossed.
For charges like those of  \cite{Penrose:1982wp}, only special deformations of the surface can be made and the connection of the charge to symmetries is lost.

Throughout our discussion, we make use of the myriad properties of CKY forms. We will denote CKY forms by $K$ and their divergence (up to a prefactor) by $\hat{K}$ (see \eqref{eq:Khat_def}). 
For the most part, we do not make use of the Einstein equations and our results apply to any geometry admitting the relevant tensors, regardless of the gravitational theory under consideration. That is, many of our results apply equally to theories of gravity coupled to matter or to  higher-derivative theories of gravity. 
In section~\ref{sec:Einstein_spaces}, however, we analyse solutions of the Einstein equation $G_{\mu\nu}+\Lambda g_{\mu\nu}=0$ in particular.
These spaces have the property that the divergence $\hat{K}$ of a CKY 2-form $K$ is a Killing vector, although this is not true in general. It is then natural to compare the charges associated with $K$ to more standard charges defined purely from $\hat{K}$. Indeed, on Einstein spaces, we find a simple relation between our currents and the Komar current.

The remainder of this paper is structured as follows. In section~\ref{sec:main_point} we review the properties of CKY forms and discuss higher-form currents defined from them. In section~\ref{sec:Einstein_spaces} we focus on 2-form currents on Einstein spaces, including a generalisation of the standard Komar currents and their relation to the currents built from CKY forms. Section~\ref{sec:examples} gives several applications of our construction, including examples of various charges which can be measured using CKY forms. We also report new CKY forms on D-brane solutions of ten-dimensional type II supergravity and discuss the associated charges. Finally, we give some concluding remarks in section~\ref{sec:conclusion}. 
Some technical details are given in the appendices.

\section{Currents and conformal Killing-Yano tensors}
\label{sec:main_point}

In this section, we consider spaces admitting CKY tensors of arbitrary rank and the conserved currents which can be constructed from them. 

\subsection{Primary and secondary currents}
\label{sec:primary_secondary}

We will be interested here in  
pairs of currents $J_{(p)}$, $J_{(p-1)}$ where $J_{(p-1)}$ is a ($p-1$)-form that is conserved
\begin{equation}
    \nabla^{\mu_1} (J_{(p-1)})_{\mu_1\dots\mu_{p-1}} = 0 \,
\end{equation}
and is
the  divergence  of the $p$-form $J_{(p)}$:
\begin{equation}
    (J_{(p-1)})_{\mu_1\dots\mu_{p-1}} = \nabla^\nu (J_{(p)})_{\nu\mu_1\dots\mu_{p-1}} \,.
\end{equation}
If  $J_{(p-1)}\ne 0$, then, following \cite{Hull:2023iny}, we refer to the lower-rank conserved current $J_{(p-1)}$ as a \emph{primary current} and the higher-rank current $J_{(p)}$ as a \emph{secondary current}.
We will be primarily concerned with the case in which both currents are globally defined (so that $J_{(p-1)}$ is co-exact) and have a local expression in terms of the curvature tensor and special global tensor fields on the spacetime.
In the special cases in which $J_{(p-1)}= 0$, then $J_{(p)}$ is itself conserved and could be referred to as a primary current.
We will often be interested in the case in which $J_{(p-1)}\ne 0$ only in certain limited regions, so that $J_{(p)}$ is conserved outside those regions.
 
A charge can be defined by integrating the Hodge dual of a primary current $J_{(p-1)}$ over a submanifold $\Sigma$ of codimension-($p-1$):
\begin{equation}\label{eq:Q}
    Q = \int_{\Sigma} \star J_{(p-1)} = \int_{\partial\Sigma} \star J_{(p)} \,.
\end{equation}
Clearly, if $\partial\Sigma=0$ (or if   $J_{(p-1)}$ vanishes) then the charge vanishes. 
The conservation of the current $J_{(p-1)}$ implies the usual conservation properties of the charge. Specifically, the charge is unchanged if $\Sigma$ is deformed to another codimension-($p-1$) submanifold with the same boundary. More generally $\Sigma$ can be deformed to a codimension-($p-1$) submanifold $\Sigma'$ without changing the charge if there exists a codimension-($p-1$) submanifold $V$ with $\int_V \star J_{(p-1)}=0$ such that $V \cup \Sigma \cup \overline{\Sigma'}$ is a closed submanifold without boundary (where the bar indicates orientation reversal).
The standard example for $p=2$ is for $\Sigma', \Sigma$ to be discs $r \le R$ in timelike hypersurfaces $t=T'$
and $t=T$, with $V$ the cylinder $r=R$, $T<t<T'$.
Such charges are often referred to as topological charges.

The $p$-form current $J_{(p)}$ is conserved in regions in which the current $J_{(p-1)}$ vanishes.
Then for $S$ a codimension-$p$ cycle in a region where $J_{(p-1)}$ vanishes there is a charge
\begin{equation}\label{eq:q}
    q = \int_{S} \star J_{(p)} 
\end{equation}
and this is unchanged under any deformation of $S$ that does not cross any region where $J_{(p-1)}\ne 0$, so that in such a region the charge depends only on the homology of $S$.
In regions where $J_{(p-1)}\neq0$, the charge \eqref{eq:q} is unchanged if $S$ is deformed to another  codimension-2 cycle $S'$ in the same homology class (i.e.~$S' = S + \partial \sigma$ with $\sigma$ a codimension-1 submanifold) if $\int_\sigma \star J_{(p-1)} = 0$. In this case the charge is not invariant under arbitrary deformations and will be called `quasi-local' or `quasi-topological'. If  $J_{(p-1)}$ vanishes everywhere, then $J_{(p)}$ is conserved everywhere and $q$ is a topological charge.

\subsection{Conformal Killing-Yano tensors}
\label{sec:CKY}

A rank-$p$ CKY tensor on a $d$-dimensional manifold $M$ is a $p$-form, $K$, which satisfies \cite{Tachibana1969OnSpace}
\begin{equation}\label{eq:CKY_eq}
    \nabla_{\nu} K_{\mu_1 \mu_2 \dots \mu_{p}} = \tilde{K}_{\nu \mu_1 \mu_2 \dots \mu_{p}} + p g_{\nu [\mu_1} \hat{K}_{\mu_2 \dots \mu_p]} \, ,
\end{equation}
where
\begin{equation}\label{eq:Khat_def}
    \tilde{K}_{\nu\mu_1\dots\mu_p} := \nabla_{[\nu} K_{\mu_1\dots\mu_p]} \qc \hat K_{\mu_2 \dots \mu_p} := \frac{1}{d - p + 1} \nabla^{\mu_1} K_{\mu_1 \mu_2 \dots \mu_{p}} \, 
\end{equation}
and $\nabla$ is the Levi-Civita connection for the metric $g_{\mu\nu}$.
From the definition \eqref{eq:Khat_def}, $\tilde{K}$ is exact and $\hat{K}$ is co-exact so that, of course, they are closed and co-closed respectively.
A Killing-Yano (KY) tensor is a CKY tensor with $\hat{K}=0$ \cite{Yano1952}, while a closed CKY tensor is one for which $\tilde{K}=0$. The KY tensors can be thought of as a higher-rank generalisation of Killing vectors and CKY tensors as a higher-rank generalisation of conformal Killing vectors.

We now discuss further properties of $\hat{K}$.
Consider, for example, the case $p=2$. In this case, it follows as an integrability condition for \eqref{eq:CKY_eq} that $\hat{K}$ satisfies\footnote{In the case where $K$ is a KY tensor, with $\hat{K}=0$, the left-hand side of this equation immediately vanishes. The right-hand side of the equation also vanishes by the integrability conditions discussed in \cite{Lindstrom:2021dpm}.}
\begin{equation}\label{eq:KV_integrability}
    {(d-2)}\nabla_{(\mu}\hat K_{\nu)} = R\indices{^\sigma_{(\mu}} K_{\nu)\sigma} \, .
\end{equation}
If the right-hand side of this equation vanishes, then $\hat{K}$ is a Killing vector, but this need not be the case in general. It is easy to show that for Einstein spaces the right-hand side does vanish \cite{Kashiwada:1968fva}, so $\hat{K}$ is a Killing vector in these cases.
Similar results are true for $p>2$ (and $d>p$). The ($p-1$)-form $\hat{K}$ would be a KY tensor if $\nabla_\mu \hat{K}_{\sigma_2\dots\sigma_p} = \nabla_{[\mu} \hat{K}_{\sigma_2\dots\sigma_p]}$.
The equation \eqref{eq:CKY_eq} has an integrability condition which has $\nabla_\mu \hat{K}_{\sigma_2\dots\sigma_p} - \nabla_{[\mu} \hat{K}_{\sigma_2\dots\sigma_p]}$ on the left hand side and various terms involving contractions of $ K$ with the full Riemann tensor on the right hand side (see appendix~\ref{app:integrability_conditions}).
From this condition, one can show that  for constant curvature spaces $\hat{K}$ is a KY $(p-1)$-form, but on a general background space this need not be the case. While most spaces do not   admit solutions to \eqref{eq:CKY_eq}, many physically relevant spaces do.\footnote{Integrability conditions for CKY forms provide  constraints on the geometries that   admit them and have been discussed in \cite{Dietz:1981, Krtous:2008tb, Batista:2014fpa}. We discuss further conditions throughout the text and in appendix~\ref{app:integrability_conditions}.}

\subsection{2-form currents}
\label{sec:2form_currents}

We now turn to the construction of currents involving CKY forms.
In the Einstein theory linearised around Minkowski spacetime there are conserved 2-form currents built from contractions of the linearised Riemann tensor and the CKY 2-forms of Minkowski space \cite{Hull:2024xgo}. We refer to the corresponding charges as Penrose charges.
In the non-linear theory, these do not generalise to covariant conserved 2-form currents in general. However, for those special spacetimes that admit a CKY 2-form $K$ we can define a 2-form current
\begin{equation}\label{eq:J2_def}
    J_{(2)}[K]_{\mu_1\mu_2} = -\frac{1}{2} \left( R_{\mu_1\mu_2\sigma_1\sigma_2} K^{\sigma_1\sigma_2} +4 R\indices{^{\sigma_1}_{[\mu_1}} K_{\mu_2]\sigma_1} + R K_{\mu_1\mu_2} \right) \, ,
\end{equation}
where $R_{\mu\nu\rho\sigma}$ is the full Riemann tensor (\emph{not} its linearisation). This can be written more compactly as
\begin{equation}
    J_{(2)}[K]_{\mu_1\mu_2} = -3 \delta_{\mu_1\mu_2 \alpha\beta}^{\nu_1\nu_2 \gamma\delta} K_{\nu_1\nu_2} R\indices{^{\alpha\beta}_{\gamma\delta}} \,,
\end{equation}
where the generalised Kronecker delta symbol is defined by $$\delta^{\alpha_1\dots\alpha_n}_{\beta_1\dots\beta_n} = \delta^{[\alpha_1}_{\beta_1} \dots \delta^{\alpha_n]}_{\beta_n}$$
for any $n$.
The divergence of $J_{(2)}[K]$ is
\begin{equation}\label{eq:divJ2=J1}
    \nabla^\mu J_{(2)}[K]_{\mu\nu} = (d-3) J_{(1)}[\hat{K}]_\nu \, ,
\end{equation}
where
\begin{equation}\label{eq:J1_def}
    J_{(1)}[\hat{K}]_\nu := G_{\nu\sigma} \hat{K}^\sigma \, .
\end{equation}
It follows from \eqref{eq:divJ2=J1} that $J_{(1)}[\hat{K}]$ is conserved:
\begin{equation}
    \nabla^\mu J_{(1)}[\hat{K}]_\mu = 0 \, .
\end{equation}
Therefore, $J_{(1)}[\hat{K}]$ is a primary current and $J_{(2)}[K]$ is the corresponding secondary current. Crucially, we note that both the primary and secondary currents are tensorial quantities.

As in subsection~\ref{sec:primary_secondary}, there is a charge
\begin{equation}\label{eq:Penrose_charge}
    Q_{\text{P}}[K] = (d-3) \int_\Sigma \star J_{(1)}[\hat{K}] \,,
\end{equation}
where $\Sigma$ is a codimension-1 submanifold.
We will refer to this charge as a \emph{Penrose charge} (indicated by the subscript `P'), as it is analogous to the construction of charges in \cite{Penrose:1982wp} (see also \cite{Hull:2024xgo}).
If $\Sigma$ has a boundary $\partial\Sigma$, then \eqref{eq:divJ2=J1} can be used to write the charge as an integral over the boundary
\begin{equation}\label{eq:Penrose_charge_boundary}
    Q_{\text{P}}[K] =  \int_{\partial\Sigma} \star J_{(2)}[K] \,.
\end{equation}

However, if $J_{(1)}[\hat{K}]$ vanishes, then $J_{(2)}[K]$ is itself a conserved current and a charge can be defined by integrating it over a codimension-2 cycle $S$ 
\begin{equation}\label{eq:non-Penrose_charge_boundary}
    q_{\text{P}}[K] =  \int_{S} \star J_{(2)}[K] \,.
\end{equation}
This is a topological charge that depends only on the homology class of $S$.

As discussed in subsection~\ref{sec:primary_secondary}, even in regions where $J_{(1)}[\hat{K}]\neq0$ the charge \eqref{eq:non-Penrose_charge_boundary} is unchanged by deformations of $S$ to another codimension codimension-2 cycle $S'=S+\partial\sigma$ provided that $\int_\sigma \star J_{(1)}[\hat{K}] = 0$. Since the charge is not invariant under arbitrary deformations in this case, it can be said to be quasi-local. We will see examples of this in section~\ref{sec:examples}.

When Einstein's equation $G_{\mu\nu}=T_{\mu\nu}$ is satisfied, the current
$ J_{(1)}[\hat{K}]_\nu = G_{\nu\sigma} \hat{K}^\sigma$ vanishes in regions where
$T_{\mu\nu}=0$ (or more generally where the components $T_{\nu\sigma} \hat{K}^\sigma=0$) so that in such regions $J_{(2)}[K] $ is conserved and the charge \eqref{eq:non-Penrose_charge_boundary} is unchanged under any deformation of the surface that stays in the region.
However, the charge  changes in general as the surface crosses any region where $T_{\mu\nu}\ne 0$ so that the charge is a measure of  the amount of energy-momentum in the region. For example, if $\hat K$ is a timelike Killing vector, this can be a measure of the mass in the region.

Let us now discuss several properties of $J_{(2)}[K]$ and $J_{(1)}[\hat{K}]$. 
\begin{itemize}
    \item Firstly, if $K$ is a KY tensor (so $\hat{K}=0$) then \eqref{eq:divJ2=J1} implies that $J_{(2)}[K]$ is conserved. This conserved 2-form current constructed from a KY tensor was first found by Kastor and Traschen \cite{Kastor2004}. The fact that it is conserved in this case can be verified simply using the Bianchi identities and defining properties of KY tensors.
    The corresponding charge is \eqref{eq:non-Penrose_charge_boundary}.
    \item Let us consider the more interesting case where $K$ is not a KY 2-form (so $\hat{K}\neq0$). Although in general it does not follow from \eqref{eq:CKY_eq} that $\hat{K}$ is a Killing vector, this is the case on constant curvature spaces and several other examples of interest (e.g.~a large family of black hole spacetimes in four dimensions \cite{Kubiznak:2007kh}). If $\hat{K}$ is a Killing vector, then $J_{(1)}[\hat{K}]_\mu = G_{\mu\nu} \hat{K}^\nu$ is the standard conserved 1-form current representing the momentum (or angular momentum) density associated with the isometry generated by $\hat{K}$. Therefore, if a spacetime admits a Killing vector which can be written as the divergence of a CKY 2-form, $J_{(2)}[K]$ gives a covariant secondary current for the primary 1-form current $J_{(1)}[\hat{K}]$ \cite{Penrose:1982wp,Goldberg:1990ju,Glass:1995pd,Glass:1998ka}. We will discuss an example of this type in detail in section~\ref{sec:examples}.
    In general, a Killing vector need not be expressible as the divergence of a CKY 2-form; for example, the axial Killing vector of the Kerr geometry cannot be written in this manner at finite $r$ \cite{Glass:1995pd}.

    \item The final case is that in which $K$ is a CKY 2-form with $\hat{K}\neq0$, but $\hat{K}$ is not a Killing vector. In this case, it still follows from \eqref{eq:divJ2=J1} that $J_{(1)}[\hat{K}]$ is conserved. If $\hat{K}$ is not a Killing vector, then the conserved current $J_{(1)}[\hat{K}]$
    gives rise to a charge which does \emph{not} stem from an isometry.
    In section~\ref{sec:examples} we will see an example of a CKY 2-form whose divergence is not a Killing vector and can still be used to define a charge in this way. Other examples of this construction have been discussed in \cite{Kinoshita:2024wyr}.
\end{itemize}

While the conservation of $J_{(1)}[\hat{K}]$ follows most directly from \eqref{eq:divJ2=J1}, it can also be derived directly. As mentioned above, this is simple in the case where $\hat{K}$ is a Killing vector, but in the more general case 
in which $\hat{K}$ is not a Killing vector
 we have
\begin{equation}
\begin{split}\label{eq:divJ1=0_explicit}
    \nabla^\mu J_{(1)}[\hat{K}]_\mu &= G_{\mu\nu} \nabla^{(\mu} \hat{K}^{\nu)} \\
    &= \frac{1}{d-2} G_{\mu\nu} R^{\sigma(\mu} K\indices{^{\nu)}_\sigma} \\
    &= \frac{1}{d-2} \left( R_{\mu\nu} R^{\sigma\mu} K\indices{^\nu_\sigma} - \frac{1}{2} R R^{\sigma\mu} K_{\mu\sigma} \right) = 0 \,,
\end{split}
\end{equation}
where we have used $\nabla^\mu G_{\mu\nu} = 0$  and  \eqref{eq:KV_integrability}. The final result vanishes since both the Ricci tensor and the contracted pair of Ricci tensors are symmetric, while $K$ is anti-symmetric.

The conservation of $J_{(1)}[\hat{K}]$ above is  non-trivial. To exemplify this, we consider a 1-form current $J_{(1)}[L]$ defined as in \eqref{eq:J1_def} where $L$ is an arbitrary 1-form. Demanding that $J_{(1)}[L]$ is conserved puts constraints on $L$. For an arbitrary 1-form $L$, we have
\begin{equation}\label{eq:J1[L]_conserved}
    \nabla^\mu J_{(1)}[L]_\mu = \nabla^\mu (G_{\mu\nu} L^\nu) = G_{\mu\nu} \nabla^{(\mu} L^{\nu)} \,,
\end{equation}
so the conservation of $J_{(1)}[L]$ is related to the properties of $\nabla^{(\mu} L^{\nu)}$. The anti-symmetrised part $\nabla^{[\mu} L^{\nu]}$ is projected out by the symmetry of the Einstein tensor.

The right-hand side of \eqref{eq:J1[L]_conserved} immediately vanishes if $L$ is a Killing vector, and in this case the conservation of the 1-form current $J_{(1)}[\hat{K}]$   corresponds simply to the conservation of the momentum (or angular momentum) associated with the isometry.
A slight generalisation is for $L$ to be a conformal Killing vector. In this case, $\nabla_{(\mu} L_{\nu)} = d^{-1} g_{\mu\nu} \nabla^\sigma L_\sigma$ so it follows from \eqref{eq:J1[L]_conserved} that $J_{(1)}[L]$ is conserved if the Ricci scalar vanishes. It is not obvious from \eqref{eq:J1[L]_conserved} that the current can be conserved for any other choices of $L$. However,  we saw above that when $L=\hat{K}$ is the divergence of a CKY 2-form, the current is conserved due to non-trivial integrability conditions and symmetry arguments. On the other hand, these more complicated arguments can be bypassed immediately once we know that $J_{(1)}[\hat{K}]$ is the divergence of a 2-form $J_{(2)}[K]$.
While this is not the most general choice of $L$ for which $J_{(1)}[L]$ is conserved, it demonstrates that studying conservation equations of the form of \eqref{eq:J1[L]_conserved} is not sufficient to determine all the conserved currents on a given spacetime as they may depend, for example, on non-trivial integrability conditions.

\subsection{Higher-rank currents}
\label{sec:higher_rank_currents}

Charges associated with 2-form currents are constructed by integrating their Hodge duals over a codimension-2 cycle, as in \eqref{eq:Penrose_charge_boundary} and \eqref{eq:non-Penrose_charge_boundary}. Within a spacelike hypersurface, a codimension-2 cycle topologically links with a point and so these integrals naturally measure the charges of point-like objects in a given hypersurface, e.g.~a massive particle or a black hole.\footnote{In the full spacetime, a codimension-2 cycle links with a line which would then naturally be interpreted as the world-line of the point-like object within a hypersurface. This is the interpretation favoured by many modern discussions of symmetry operators (e.g.~\cite{Gaiotto:2014kfa}), but is clearer in Euclidean space.} It is then natural to ask whether charges carried by extended objects (e.g.~black strings or brane-like objects) can be measured by an integral defined on a higher codimension surface. Such charges can be found by integrating the Hodge duals of higher-rank currents. We now analyse such higher-rank currents which are analogous to the 2-form currents $J_{(2)}[K]$ introduced in \eqref{eq:J2_def}.

Given a CKY $p$-form $K$, we construct the following $p$-form current
\begin{equation}\label{eq:Jn_def}
    J_{(p)}[K]_{\mu_1\dots\mu_p} = -\frac{(p+2)(p+1)}{4} \delta^{\nu_1\dots\nu_p\gamma\delta}_{\mu_1\dots\mu_p\alpha\beta} K_{\nu_1\dots\nu_p} R\indices{^{\alpha\beta}_{\gamma\delta}} \, ,
\end{equation}
for $p\geq 1$. 
We reserve the letter $K$ for CKY tensors, while their curl and divergence are denoted by $\tilde{K}$ and $\hat{K}$ as defined in \eqref{eq:Khat_def}. 
Expanding, for $p>1$ we find
\begin{equation}\label{eq:Jn_expanded}
    J_{(p)}[K]_{\mu_1\dots\mu_p} = -\frac{p(p-1)}{4} R\indices{_{[\mu_1\mu_2}^{\rho\sigma}} K_{\mu_3\dots\mu_p]\rho\sigma} + p (-1)^{p-1} R\indices{^\rho_{[\mu_1}} K_{\mu_2\dots\mu_p]\rho} -\frac{1}{2} R K_{\mu_1\dots\mu_p} \, ,
\end{equation}
which coincides with \eqref{eq:J2_def} when $p=2$. For $p=1$ the definition \eqref{eq:Jn_def} reproduces \eqref{eq:J1_def}.
Taking a covariant divergence, we find
\begin{equation}\label{eq:divJn}
    \nabla^{\mu_1}J_{(p)}[K]_{\mu_1\dots\mu_p} = (d-p-1) J_{(p-1)}[\hat{K}]_{\mu_2\dots\mu_p} \, ,
\end{equation}
where we have used the CKY equation \eqref{eq:CKY_eq} as well as the Bianchi identities $\nabla_{[\mu}R_{\rho\sigma]\kappa\lambda}=0$ and $R_{[\mu\nu\rho]\sigma} = 0$. This reduces to \eqref{eq:divJ2=J1} for $p=2$.
Taking a further divergence of \eqref{eq:divJn} implies that $J_{(p-1)}[\hat{K}]$ is conserved (provided $p\neq d-1$), that is,
\begin{equation}\label{eq:J[Khat]_conserved}
    \nabla^{\mu_2} J_{(p-1)}[\hat{K}]_{\mu_2\mu_3\dots\mu_p} = 0 \, .
\end{equation}
Therefore, $J_{(p-1)}[\hat{K}]$ is a primary current and $J_{(p)}[K]$ is the corresponding secondary current. As for the $p=2$ case in the previous subsection, both the primary and secondary currents are covariant.

Equations~\eqref{eq:divJn} and \eqref{eq:J[Khat]_conserved} are the main results of the this work. They describe a `nested' structure for the currents \eqref{eq:Jn_def}: taking the divergence of the rank-$p$ current with an argument given by a CKY $p$-form $K$ results in the rank-$(p-1)$ current with an argument given by the divergence of $K$ (up to a prefactor).  Equation~\eqref{eq:J[Khat]_conserved} then shows that the lower-rank current is conserved. Therefore, given a CKY $p$-form we can immediately construct both a conserved $(p-1)$-form primary current as well as a $p$-form secondary current for it. This relation of currents then implies that the associated charges can be directly related, in a manner analogous to \eqref{eq:Penrose_charge} for the case $p=2$. 
The charge $Q_{\text{P}}[K] $ constructed by integrating $ J_{(p-1)}[\hat{K}]$ 
over  a codimension-$(p-1)$ submanifold $\Sigma$ can be re-expressed as an integral of $J_{(p)}[K] $ over the boundary of $\Sigma$:
\begin{equation}
    Q_{\text{P}}[K] = (d-p-1)\int_{\Sigma} \star J_{(p-1)}[\hat{K}] = \int_{\partial\Sigma} \star J_{(p)}[K] \,.
\end{equation}
As for the $p=2$ case described in the previous subsection, we will refer to these quantities as Penrose charges.

If $J_{(p-1)}[\hat{K}]$ vanishes, then $J_{(p)}[K]$ is itself a conserved current.
This will be the case if $K$ is a KY tensor so that $\hat{K}=0$, in which case $J_{(p)}[K]$ has been studied in \cite{Kastor2004}.
Then a topological charge can be defined by integrating this over a codimension-$p$ submanifold $S$,
\begin{equation}\label{eq:non-Penrose_charge_boundary_p}
    q_{\text{P}}[K] =  \int_{S} \star J_{(p)}[K] \,,
\end{equation}
that depends only on the homology class of $S$.

We now discuss several properties of the currents $J_{(p)}[K]$ and $J_{(p-1)}[\hat{K}]$ that are analogous to the case $p=2$ discussed in the previous subsection. 
\begin{itemize}
    \item Firstly, consider the case where $K$ is a KY $p$-form, i.e.~a CKY $p$-form with $\hat{K}=0$. Then \eqref{eq:divJn} implies that $J_{(p)}[K]$ is conserved. 
    The conservation can be shown directly using only Bianchi identities. 

    \item In the more general case where $\hat{K}$ is non-zero, then \eqref{eq:J[Khat]_conserved}   implies that $J_{(p-1)}[\hat{K}]$ is conserved. In some cases (e.g.~on maximally symmetric spaces) $\hat{K}$ is itself a KY $(p-1)$-form. Then, from \eqref{eq:divJn}, $J_{(p)}[K]$ gives a secondary current for $J_{(p-1)}[\hat{K}]$ related to the KY tensor. 

    \item Finally, there is the interesting possibility of having a CKY $p$-form $K$ whose divergence $\hat{K}$ is not a KY $(p-1)$-form. In this case, \eqref{eq:J[Khat]_conserved} still implies that $J_{(p-1)}[\hat{K}]$ is conserved, but it does not relate to a KY tensor.
    The functional form of $J_{(p-1)}[\hat{K}]$ is still the same as the current studied in \cite{Kastor2004}, but we see that more general tensors than   KY forms can be used to define conserved currents.     We will see examples of this type in section~\ref{sec:examples}.
\end{itemize}

\subsection{Integrability conditions and conservation}

If $\hat{K}$ is a KY tensor, the fact that $J_{(p-1)}[\hat{K}]$ is conserved can be shown directly using only the Bianchi identities. Let us consider the case where $\hat{K}$ is not a KY tensor. In this case, it is complicated to show explicitly that $J_{(p-1)}[\hat{K}]$ is conserved. The quickest way is to note that it can be written as the divergence of the $p$-form $J_{(p)}[K]$ as in \eqref{eq:divJn}, but this is somewhat indirect. In the case of $p=2$, we showed in \eqref{eq:divJ1=0_explicit} that $J_{(1)}[\hat{K}]$ is conserved even when $\hat{K}$ is not a Killing vector by using integrability conditions of the CKY equation and symmetry arguments. The same is possible, though far less obvious, in the higher-rank case. Let us take the first non-trivial example of $p=3$, so $\hat{K}$ is a 2-form. Explicitly taking the divergence of $J_{(2)}[\hat{K}]$ as defined in \eqref{eq:Jn_def} we find, after a long sequence of manipulations outlined in appendix~\ref{app:integrability_2nd_rank},
\begin{equation}\label{eq:divJ2_integrability}
    {4(d-3)}\nabla_\mu J_{(2)}[\hat{K}]^{\mu\nu} =- R\indices{^{\rho\sigma \nu}_{[\alpha}} R_{\beta\gamma]\rho\sigma} K^{\alpha\beta\gamma} \, .
\end{equation}
Since $J_{(2)}[\hat{K}]$ is conserved from \eqref{eq:J[Khat]_conserved}, the right-hand side of \eqref{eq:divJ2_integrability} must vanish. This could be seen as a constraint on the Riemann tensor of a geometry if it is to admit the CKY tensor $K$.
Here we have derived this constraint using \eqref{eq:J[Khat]_conserved}. In appendix~\ref{app:integrability_2nd_rank} we verify directly that the left-hand side of \eqref{eq:divJ2_integrability} vanishes using only integrability conditions of the CKY equation (i.e.\ without using the fact that $J_{(2)}[\hat{K}]$ can be written as the divergence of a 3-form), thereby proving that the composite quantity on the right-hand side does, indeed, vanish. This is more direct but considerably more complicated. To the authors' knowledge this is a new relation, and integrability conditions of this sort for CKY tensors have not been reported previously.\footnote{See, e.g.~\cite{Tachibana1969OnSpace,Kashiwada:1968fva,Batista:2014fpa} for other integrability conditions on CKY tensors.}

It is somewhat surprising that a derivation of \eqref{eq:J[Khat]_conserved} using the only the basic CKY conditions (i.e.~without finding the secondary current $J_{(p)}[K]$ in \eqref{eq:divJn}) is quite involved.
Let us contrast this with the case of $p=2$. Then $\hat{K}$ is a 1-form and the conservation of $J_{(1)}[\hat{K}]$ was shown explicitly in \eqref{eq:divJ1=0_explicit}. The similarity with the $p=3$ case discussed here is that the result in the $p=2$ case also vanishes because of symmetry properties of the Ricci and Einstein tensors, and not solely because of properties of $K$.

For higher-rank CKY tensors the same pattern continues. Namely, explicitly evaluating the divergence of $J_{(p-1)}[\hat{K}]$, which we know to vanish from \eqref{eq:J[Khat]_conserved}, gives a condition on the Riemann tensor and $K$. For $p>2$, this condition is
\begin{equation}
    R^{\kappa\lambda\alpha\beta} K\indices{^{\sigma[\mu_2\dots\mu_{p-2}}_{\kappa\lambda}} R\indices{^{\mu_{p-1}]}_{\sigma\alpha\beta}} = 0 \, .
\end{equation}
We expect that it is possible to show this using integrability conditions of the CKY equation, in analogy to the $p=3$ case discussed in appendix~\ref{app:integrability_2nd_rank}, but the derivation presented above starting from the secondary current $J_{(p)}[K]$ is much more streamlined.

It is clear from \eqref{eq:divJn} that $p=d-1$ is a special case. For this rank, the right-hand side of \eqref{eq:divJn} vanishes and we cannot infer that $J_{(p-1)}[\hat{K}]$ is conserved.
Rather, in this case \eqref{eq:divJn} shows that $J_{(p)}[K]$ is conserved identically. In fact, this was guaranteed since it is known that the current $J_{(p)}[K]$ vanishes when $p=d-1$ regardless of the constraints on $K$ \cite{Lindstrom:2021qrk}.\footnote{This can be seen by writing the $J_{(p)}[K]$ in terms of the Weyl tensor and using the fact that the Weyl tensor is traceless.}
This fact has important implications for the co-exactness of rank-$(d-2)$ conserved currents.
For example, let us consider $p=3$, where $K$ is a rank-3 CKY tensor and its divergence $\hat{K}$ is a 2-form. The discussion above leads us to consider the 2-form current $J_{(2)}[\hat{K}]$. In $d>4$ dimensions, $J_{(3)}[K]$ is a secondary current for $J_{(2)}[\hat{K}]$ from \eqref{eq:divJn}. However, in $d=4$ dimensions, $J_{(3)}[K]$ (and, therefore, its divergence) vanishes identically and so $J_{(2)}[\hat{K}]$ is no longer co-exact. It is then not guaranteed that $J_{(2)}[\hat{K}]$ is conserved. In cases where $\hat{K}$ is a KY tensor, however, it remains true that $J_{(2)}[\hat{K}]$ is co-closed.
This is also true for linearisations of these currents around some fixed background and has been important in determining the non-trivial higher-form symmetries of linear gravity in \cite{Benedetti:2023ipt, Hull:2024xgo}.

Let us elaborate on the non-trivial manner in which the currents discussed above are conserved. We will focus on the case where $p=3$, but a similar discussion applies for any $p\geq 3$. Consider a 2-form current $J_{(2)}[M]$ as defined in \eqref{eq:Jn_def} where $M$ is an arbitrary 2-form, 
\begin{equation}
    J_{(2)}[M]_{\mu_1\mu_2} = - 3 \delta^{\nu_1\nu_2\gamma\delta}_{\mu_1\mu_2\alpha\beta} M_{\nu_1\nu_2} R\indices{^{\alpha\beta}_{\gamma\delta}} \, .
\end{equation}
Demanding that $J_{(2)}[M]$ is conserved puts constraints on $M$. The divergence of the previous equation gives, using the Bianchi identity $\nabla_{[\mu} R_{\alpha\beta]\gamma\delta}=0$,
\begin{equation}\label{eq:divJ[L]_working}
    \nabla^{\mu_1} J_{(2)}[M]_{\mu_1\mu_2} = -3 \delta_{\mu_1\mu_2\alpha\beta}^{\nu_1\nu_2\gamma\delta} R\indices{^{\alpha\beta}_{\gamma\delta}} \nabla^{\mu_1} M_{\nu_1\nu_2} \, .
\end{equation}
The conservation of $J_{(2)}[M]$ is thus governed by the properties of the tensor $\nabla_{\mu_1} M_{\nu_1\nu_2}$.
The simplest case is to consider $\nabla_{\mu_1} M_{\nu_1\nu_2} = \nabla_{[\mu_1} M_{\nu_1\nu_2]}$, for which the result vanishes using a Bianchi identity $R_{[\mu\nu\rho]\sigma}=0$ after some short manipulations. In this case $M$ is a KY 2-form. This is analogous to the simplest case discussed under \eqref{eq:J1[L]_conserved} for $p=2$, where $L$ is a Killing vector. 
A possible generalisation is for $M$ to be a CKY 2-form (satisfying \eqref{eq:CKY_eq}) but in this case the result only vanishes on a Ricci flat geometry. This is analogous to the case where $L$ is a conformal Killing vector in the $p=2$ discussion under \eqref{eq:J1[L]_conserved}. 

Above, we have seen that when $M=\hat{K}$ is the divergence of a CKY 3-form, the current is also conserved due to very non-trivial integrability conditions and symmetry arguments. As discussed under \eqref{eq:J1[L]_conserved} for $p=2$, we see that studying conservation equations of the form \eqref{eq:divJ[L]_working} is not sufficient to determine all the conserved quantities on a given spacetime since non-trivial integrability conditions can be essential in determining under which conditions the right-hand side vanishes. 

For higher-rank currents, the same pattern continues. The simplest conserved current of the form $J_{(p-1)}[M]$ for a $(p-1)$-form $M$ is when $M$ is a KY $(p-1)$-form. In this case, the covariant derivative $\nabla M$ is fully anti-symmetric and the current is conserved from a simple Bianchi identity.
If the spacetime admits more structure then $\nabla M$ can take a more general form and still yield a conserved current. In the case where there exists a rank-$p$ CKY tensor $K$, then taking $M=\hat{K}$ gives a current which is conserved owing to non-trivial integrability conditions and symmetry arguments.

\section{Currents and charges on Einstein spaces}
\label{sec:Einstein_spaces}

In the previous section, we have discussed the conservation of $J_{(p-1)}[\hat{K}]$ on general spacetimes. In this section, we consider these currents with $p=2$ on Einstein spacetimes admitting CKY 2-forms, where some additional properties can be derived with clear application to physically relevant spacetimes in General Relativity.\footnote{Other non-generic situations where currents involving (C)KY forms have extra properties have been studied, e.g.~in \cite{Lindstrom:2021qrk,Lindstrom:2021dpm}.} We also discuss Komar integrals, generalisations thereof, and their connection to the currents $J_{(p)}[K]$ on spacetimes admitting CKY 2-forms.\footnote{Other generalisations of the Komar 2-form and their applications to Smarr formulae have been discussed recently in \cite{Bandos:2024pns} and references therein.}

\subsection{Conservation of \texorpdfstring{$J_{(2)}[K]$}{J2[K]} on Einstein spaces}

Recall from \eqref{eq:divJ2=J1} that there is a conserved 1-form current $J_{(1)}[\hat{K}]$, for which $J_{(2)}[K]$ gives a secondary current. The conservation of the 1-form current leads to the Penrose charge in \eqref{eq:Penrose_charge}. It can be written as a surface integral of $J_{(2)}[K]$ as in \eqref{eq:Penrose_charge_boundary}. We now investigate the cases where the 2-form current $J_{(2)}[K]$ (or some suitable generalisation thereof) is itself conserved.

Indeed, on general spacetimes it follows from \eqref{eq:divJ2=J1} that $J_{(2)}[K]$ is not conserved. On a Ricci flat spacetime, however, $J_{(1)}[\hat{K}]=0$ so that the Penrose charge  \eqref{eq:Penrose_charge} is zero and
$J_{(2)}[K]$ is conserved.\footnote{We note that this is a property of the rank-2 current only. The $J_{(p)}[K]$ with $p>2$ are not generically conserved on Ricci flat spaces since $J_{(p-1)}[\hat{K}]$ depends on the full Riemann tensor, rather than just the Ricci tensor, and so the right-hand side of \eqref{eq:divJn} does not generally vanish.} 
In this case, the conserved current $J_{(2)}[K]$ can be integrated over a codimension-2 cycle $S$ to construct the topological charge $q_{\text{P}}[K]$ in \eqref{eq:non-Penrose_charge_boundary}.

This result applies only to Ricci flat spaces but can be generalised to Einstein spaces as follows. An Einstein spacetime is one satisfying 
\begin{equation}\label{eq:Einstein}
    G_{\mu\nu} + \Lambda g_{\mu\nu}=0
\end{equation}
for some constant $\Lambda$. We recall from section~\ref{sec:CKY} that on Einstein spaces the divergence $\hat{K}$ of a CKY 2-form $K$ is a Killing vector, which follows from \eqref{eq:KV_integrability}.
It is possible to modify $J_{(2)}[K]$ to obtain a current which is conserved on any Einstein space. Specifically, we define
\begin{equation}\label{eq:JLambda}
    J_{(2)}^\Lambda[K]_{\mu\nu} = J_{(2)}[K]_{\mu\nu} + \frac{d-3}{d-1} \Lambda K_{\mu\nu}\,,
\end{equation}
which, using \eqref{eq:Einstein}, satisfies $\nabla^\mu J_{(2)}^\Lambda[K]_{\mu\nu}=0$. Therefore, on Einstein spaces, we can construct conserved 2-form currents associated with all CKY 2-forms. Integrating the current gives a charge
\begin{equation}\label{eq:Penrose_charge_Lambda}
    q_{\text{P}}^\Lambda[K] = \int_S \star J_{(2)}^\Lambda[K] 
\end{equation}
which reduces to $q_{\text{P}}[K]$ when $\Lambda=0$.
We note that on an Einstein space, using \eqref{eq:J2_def}, $J^\Lambda_{(2)}[K]$ can be written
\begin{equation}\label{eq:Penrose_Lambda_simpler}
    J_{(2)}^\Lambda[K]_{\mu\nu} = -\frac{1}{2} R_{\mu\nu\alpha\beta} K^{\alpha\beta} + \frac{2}{(d-1)(d-2)} \Lambda K_{\mu\nu} \, .
\end{equation}

\subsection{Komar charges and generalisations}
\label{sec:Komar_charges}

In this subsection we first review the standard construction of Komar charges on spacetimes with isometries, and then generalise this construction slightly to Einstein spacetimes admitting a CKY 2-form. In the following subsection, we show that these charges can be related to those discussed in the previous subsection.

For a spacetime admitting a Killing vector $k$, the 1-form current
\begin{equation}
    f_{(1)}[k]_\mu = R_{\mu\nu} k^\nu
\end{equation}
is conserved. This is most simply shown by observing that it is proportional to the divergence of a 2-form. Namely, the Komar 2-form 
\begin{equation}\label{eq:Komar_2_form_def}
    F_{(2)}[k]_{\mu\nu} = -\nabla_{[\mu} k_{\nu]}
\end{equation}
has the property that
\begin{equation}\label{eq:Komar_working}
    \nabla^\mu F_{(2)}[k]_{\mu\nu} = R_{\nu\rho} k^\rho = f_{(1)}[k]_\nu \, ,
\end{equation}
where the Ricci identity has been used. It immediately follows that $f_{(1)}[k]$ is conserved. The integral of $\star f_{(1)}[k]$ over a $(d-1)$-volume $\Sigma$ then gives the Komar charge which, from \eqref{eq:Komar_working}, can be written as a surface integral of $\star F_{(2)}[k]$:
\begin{equation}\label{eq:Komar_charge}
    Q_{\text{K}}[k] = \int_\Sigma \star f_{(1)}[k] = \int_{\partial\Sigma} \star F_{(2)}[k] \,.
\end{equation}

In the previous subsection, we found a generalisation of $J_{(2)}[K]$ which is itself conserved on an Einstein space. Let us now discuss the analogous modification of the Komar 2-form $F_{(2)}[k]$. As for $J_{(2)}[K]$, the Komar 2-form $F_{(2)}[k]$ is conserved on Ricci flat spacetimes from \eqref{eq:Komar_working}, since in this case $f_{(1)}[k]=0$, and can be integrated over a closed codimension-2 submanifold $S$ to give a charge
\begin{equation}\label{eq:Komar_charge_boundary}
    q_{\text{K}}[k] = \int_S \star F_{(2)}[k]
\end{equation}
which only depends on the homology class of $S$.
Let us now restrict to the case where the Killing vector $k$ can be written as the divergence of a CKY 2-form $K$, i.e.~$k=\hat{K}$.\footnote{While it is true that, on an Einstein space, the divergence of a CKY 2-form gives a Killing vector (see section~\ref{sec:CKY}), the converse is not true in general. For example, the axial Killing vectors of the Schwarzschild and Kerr spacetimes cannot be written in this manner \cite{Glass:1995pd}.}
We define the generalised Komar 2-form
\begin{equation}\label{eq:KomarLambda}
    F_{(2)}^\Lambda [K]_{\mu\nu} = F_{(2)}[\hat{K}]_{\mu\nu} - \frac{2}{(d-1)(d-2)} \Lambda K_{\mu\nu} \,.
\end{equation}
By similar manipulations to \eqref{eq:Komar_working}, this current is conserved on Einstein spaces. Note that this generalisation to non-Ricci flat spacetimes is not possible if the Killing vector $k$ cannot be written as the divergence of a CKY 2-form. A charge can be defined by
\begin{equation}\label{eq:Komar_charge_Lambda}
    q_{\text{K}}^\Lambda[K] = \int_S \star F_{(2)}^\Lambda[K] \, .
\end{equation}
In the case where $\Lambda=0$ this reduces to the standard Komar charge $q_{\text{K}}[k]$. We will refer to $q_{\text{K}}^\Lambda[K]$ as a \emph{generalised Komar charge}.

\subsection{Comparing charges: Komar vs.~Penrose}
\label{sec:comparing_charges}

In this section, we have discussed a pair of 1-form currents $J_{(1)}[\hat{K}]$ and $f_{(1)}[k]$, and their respective 2-form secondary currents $J_{(2)}[K]$ and $F_{(2)}[k]$. In the case where $\hat{K}$ is a Killing vector, i.e.~$k=\hat{K}$, it is natural to compare the two.
On a general spacetime with an isometry, the 1-form currents $J_{(1)}[\hat{K}]_\mu = G_{\mu\nu}\hat{K}^\nu$ with $k=\hat{K}$ and $f_{(1)}[k]_\mu = R_{\mu\nu}k^\nu$ are both conserved and differ by a third current
\begin{equation}
    V_\mu = \frac{1}{2} R k_{\mu}
\end{equation}
which is also conserved.\footnote{The conservation of $V_\mu$ can also be seen from the fact that $k^\mu \nabla_\mu R=0$ for a Killing vector $k$.}
The two currents $J_{(1)}[k]_\mu $ and $f_{(1)}[k]_\mu$ then only agree if the difference $V_\mu$ vanishes, which will be the case for spacetimes for which the Ricci scalar vanishes, $R=0$.
Note that if the Ricci scalar is constant then the current $V_\mu$ can be written as
\begin{equation}
    V_\mu = \frac{1}{2} \nabla^\nu (R K_{\mu \nu})
\end{equation}
and the charge given by integrating $\star V$ over a region of codimension 1 can be rewritten as a surface integral over the boundary of the region.
In particular, this will be the case for the Einstein spaces considered in the next subsection.

On an Einstein space admitting a CKY 2-form $K$, we have two charges, $q_{\text{P}}^\Lambda[K]$ in \eqref{eq:Penrose_charge_Lambda} and $q_{\text{K}}^\Lambda[K]$ in \eqref{eq:Komar_charge_Lambda}, which can be related. An integrability condition for \eqref{eq:CKY_eq} is that
\begin{equation}
    \frac{1}{2} R_{\mu\nu\rho\sigma} K^{\rho\sigma} = R_{\sigma \mu} K\indices{^\sigma_\nu} + \nabla^\sigma \tilde{K}_{\sigma\mu\nu} + (d-2) \nabla_\mu \hat{K}_\nu \, ,
\end{equation}
where $\tilde{K}$ is defined in \eqref{eq:Khat_def}.
Symmetrising this equation over the $\mu$, $\nu$ indices recovers \eqref{eq:KV_integrability}. Anti-symmetrising this equation and using \eqref{eq:Einstein} leads, after some simple calculation, to
\begin{equation}\label{eq:Komar=KTcurrents}
    J_{(2)}^\Lambda[K]_{\mu\nu} = (d-2) F_{(2)}^\Lambda[K]_{\mu\nu} - \nabla^\sigma \tilde{K}_{\sigma\mu\nu} \, .
\end{equation}
The relevant manipulations are most easily done  using the form \eqref{eq:Penrose_Lambda_simpler} of $J_{(2)}^\Lambda[K]$.
We see that the 2-form $J_{(2)}^\Lambda[K]$ and the generalised Komar 2-form $F_{(2)}^\Lambda[k]$ differ by a co-exact term. Therefore, the generalised charges are related by
\begin{equation}\label{eq:Komar=KT}
    q_{\text{P}}^\Lambda[K] = (d-2) q^\Lambda_{\text{K}}[K] \, .
\end{equation}
We stress that this result only holds on Einstein spaces, since \eqref{eq:Komar=KTcurrents} used \eqref{eq:Einstein}. In particular, for $d=4$ the Penrose charge is twice the Komar charge.

When $k$ generates translations in time, the charge $q_{\text{K}}^\Lambda[K]$ gives a measure of the energy contained within the surface $S$ referred to as the Komar mass. In the case where $k$ can be written as the divergence of a CKY 2-form $K$, \eqref{eq:Komar=KT} implies that the Komar mass is equivalently given by the charge $q^\Lambda_{\text{P}}[K]$ (up to a prefactor).

\subsection{Spacetime and matter}

Throughout this section, we have restricted attention to Einstein spaces, satisfying \eqref{eq:Einstein}. However, the results established here also have applications to spaces satisfying the Einstein equations
\begin{equation}\label{eq:einstein_source}
    G_{\mu\nu} + \Lambda g_{\mu\nu} = T_{\mu\nu} \,.
\end{equation}
The currents $J_{(2)}^\Lambda[K]$ and $F_{(2)}^\Lambda[K]$ are then only conserved in regions where $T_{\mu\nu}=0$. The associated charges $q_{\text{P}}^\Lambda [K]$ and $q_{\text{K}}^\Lambda [K]$ can be defined as integrals over a surface $S$ contained in such a region and will be invariant under deformations of $S$ that stay within the region in which $T_{\mu\nu}=0$. However, if it is deformed through a region in which $T_{\mu\nu}\neq 0$ then it will change.

The corresponding conserved primary 1-form currents are $J_{(1)}^\Lambda[K] = \dd^\dagger J_{(2)}^\Lambda[K]$ and
$f_{(1)}^\Lambda[K] = \dd^\dagger F_{(2)}^\Lambda[K]$ which, using \eqref{eq:einstein_source}, evaluate to
\begin{equation}
    J_{(1)}^\Lambda [K]_\nu = (d-3) T_{\nu\sigma} \hat{K}^\sigma \qc f_{(1)}^\Lambda [K]_\nu = \left( T_{\nu\sigma} - \frac{T}{d-2} g_{\nu\sigma} \right) \hat{K}^\sigma \,.
\end{equation}
Then if the surface $S$ bounds a volume $\Sigma$, the charges can be written as 
\begin{equation}
    q_{\text{P}}^\Lambda [K]=\int_\Sigma \star J_{(1)}^\Lambda[K] \qc q_{\text{K}}^\Lambda [K]= \int_\Sigma \star f_{(1)}^\Lambda[K]
\end{equation}
so that these are the total charges within $S$ from the charge densities $J_{(1)}^\Lambda[K]$, $f_{(1)}^\Lambda[K]$.

\section{Applications}
\label{sec:examples}

In this section, we study several applications of the constructions of the previous sections.
There are many spacetimes of physical interest which admit CKY tensors and so have conserved currents.
We will give these explicitly for a number of examples and integrate them over suitable submanifolds to find charges. Interpreting the result of this integral as a conserved quantity then relies on some global notion of time which, for general curved geometries, need not exist.
In particular, we study the Kerr-Newman black hole, the AdS-Kerr black hole, and several D-brane solutions of 10-dimensional supergravity. 

There are spaces which admit CKY tensors but do not have a global notion of time, or for which there is not a suitable submanifold over which to integrate.
For example, the Lorentzian Taub-NUT solution does not have a global time; a recent discussion of Komar charges for this space can be found in \cite{Barbagallo:2025tjr}.
The Kaluza-Klein monopole \cite{Gross:1983hb} is a product of a global time with Euclidean Taub-NUT space \cite{Gibbons:1987sp, vanHolten:1994ta}. It admits a CKY 2-form and so there is a conserved 2-form current, but there are no non-trivial 2-cycles over which to integrate it since $H_2(M)=0$ in this case.\footnote{In \cite{Jezierski:2006fw}, the integral over the base of the Hopf fibration at fixed radius was considered, but the interpretation of this is not clear.}

\subsection{Kerr-Newman black hole}
\label{sec:Kerr-Newman}

In this subsection we study the example the Kerr-Newman black hole solution.\footnote{For a review of the properties of the Kerr-Newman spacetime, see \cite{Adamo:2014baa}.} This space has a timelike Killing vector which can be written as the divergence of a CKY 2-form and so the results developed in section~\ref{sec:2form_currents} can be used to write the 1-form current $J_{(1)}[\hat{K}]$ as the divergence of the 2-form current as in \eqref{eq:divJ2=J1}. The charge contained inside the surface $r=R_0$ can then be written as a surface integral and evaluated.

In Boyer-Lindquist coordinates, the Kerr-Newman spacetime has the metric 
\begin{equation}\label{eq:KN_metric}
    \dd{s}^2 = -\frac{\Delta}{\rho^2} \left(\dd{t}-a \sin^2\theta \dd{\phi}\right)^2 + \frac{\rho^2}{\Delta(r)} \dd{r}^2 + \rho^2 \dd{\theta}^2 + \frac{\sin^2\theta}{\rho^2}\left( a \dd{t} - (r^2+a^2)\dd{\phi}\right)^2 \,,
\end{equation}
where
\begin{equation}\label{eq:KN_defs}
    \Delta = r^2 -2mr + a^2 + e^2\qc 
    \rho^2 = r^2 + a^2 \cos^2 \theta \, .
\end{equation}
The solution also has an electromagnetic potential
\begin{equation}
    A = \frac{er}{\rho^2} (\dd{t} - a \sin^2\theta \dd\phi) \,.
\end{equation}
The metric \eqref{eq:KN_metric} solves the Einstein-Maxwell equations. The Kerr-Newman metric reduces to the Reissner-Nordström solution when $a=0$, to the Kerr solution when $e=0$, and to the Schwarzschild solution when both $a=e=0$. 

The metric \eqref{eq:KN_metric} admits a timelike Killing vector 
\begin{equation}
    k = \partial_t \, ,
\end{equation}
so we can construct the conserved 1-form current $J_{(1)}[k]$ defined in \eqref{eq:J1_def}. This current density can then be integrated over a 3-volume at constant time to give a total charge in that region, which gives a measure of the energy within that region. The singular geometry means that one cannot take the 3-volume to be the interior of the surface $r=R_0$ in a constant time hypersurface (with $R_0$ outside the outer horizon of the black hole). Instead, one can integrate over a 3-volume bounded between $r=R_0$ and $r=R_1$ to give a measure of the energy between those two radial distances. Performing this integral explicitly is complicated by the form of the metric components. In \cite{Lindstrom:2022qjx, Lindstrom:2021dpm}, for example, it could not be evaluated fully.

The results of section~\ref{sec:main_point} allow us to do more.
Firstly, we note that the Kerr-Newman geometry admits a rank-2 closed CKY tensor \cite{Lindstrom:2022qjx}
\begin{equation}\label{eq:KN_CCKY}
    K = -r \dd{r} \wedge \left( \dd{t} - a \sin^2 \theta \dd{\phi} \right) - a\cos\theta \sin\theta \dd{\theta}\wedge \left( a \dd{t} - (r^2+a^2) \dd{\phi} \right) 
\end{equation}
and that this CKY has the property that its divergence gives the timelike Killing vector $k$,
\begin{equation}\label{eq:KN_KV}
    \hat{K}^\mu = k^\mu \, .
\end{equation}
Therefore, from \eqref{eq:divJ2=J1}, the 1-form current can be written as the divergence of $J_{(2)}[K]$ defined in \eqref{eq:J2_def}.
A simple application of this result is that the energy located between the radial distances $R_0$ and $R_1$ discussed above can now be written as the difference of the surface integral of $J_{(2)}[K]$ over a 2-sphere at $r=R_0$ and $r=R_1$, which is straightforward to evaluate.

Our surface integral charges, however, can be viewed as giving a measure of the energy contained within a radial distance $r=R_0$. We denote by $S_t$ the surface at fixed $r$ and $t$, with $r=R_0$ outside the horizon, which is topologically a 2-sphere.
If $e=0$, we have the Kerr solution which is Ricci flat and the 2-form current $J_{(2)}[K]$ is conserved and can be integrated over $S_t$ to give the charge \eqref{eq:non-Penrose_charge_boundary}. As $J_{(2)}[K]$ is co-closed, this charge is independent of the time $t$.

Consider now the case in which $e\ne0$ so that the geometry is not Ricci flat, the 1-form current $J_{(1)}[k]\neq0$ and so the 2-form current $J_{(2)}[K]$ is not co-closed. In this case,
$q_{\text{P}}[K]$ evaluated on a 2-sphere $S_t$ enclosing the black hole within a constant time hypersurface still leads to a charge which is conserved in time. This can be seen as follows. The difference between the charge evaluated on $S_t$ and $S_{t'}$ is
\begin{equation}\label{eq:KN_working}
    q_{\text{P}}[K,t'] - q_{\text{P}}[K,t] = \int_{S_{t'}-S_t} \star J_{(2)}[K] = \int_W \star J_{(1)}[\hat{K}] \,,
\end{equation}
where $W = S_t \times [t,t']$ is a 3-dimensional cylinder. While $J_{(1)}[\hat{K}]\neq0$, one can explicitly verify that the radial component does vanish: $J_{(1)}[\hat{K}]_r = 0$. This implies that the right-hand side of \eqref{eq:KN_working} vanishes and the charge $q_{\text{P}}[K]$ is conserved in time. This demonstrates the the charge $q_{\text{P}}[K]$ is unchanged when the surface on which it is defined is moved in the $t$ direction. This is not true if the surface is moved in the $r$ direction since $J_{(1)}[\hat{K}]_t\neq0$. Since the charge is invariant under only some deformations, it could be called quasi-local.

The charge can be explicitly evaluated with computer mathematics software, giving the result
\begin{equation}
\begin{split}\label{eq:KN_charge}
    \frac{1}{8\pi} q_{\text{P}}[K] &= -\frac{1}{2\pi}\int_0^{2\pi} \dd\phi \int_0^\pi \dd\theta \frac{(a^2+R_0^2)(a^2 m + R_0(e^2-2mR_0)+ a^2 m\cos 2\theta)\sin\theta}{(a^2+2R_0^2+a^2\cos^2 2\theta)^2} \\
    &= m - \frac{e^2}{4R_0} - \frac{e^2(a^2+R_0^2)}{4aR_0^2} \arctan (\frac{a}{R_0}) \,.
\end{split}
\end{equation}
Clearly, the $R_0$ dependence shows that energy is not localised within the horizon when the black hole is charged. This was, of course, to be expected as the electromagnetic field outside the horizon carries energy.

The result \eqref{eq:KN_charge} agrees with those presented in \cite{Aguirregabiria:1995qz} and \cite{Xulu:2003rd} where various different definitions of the energy were compared. 
Many of those definitions suffer from the limitation that the result can only be interpreted as the energy contained in a given region when the relevant integral is evaluated in asymptotically Cartesian coordinates. While \cite{Aguirregabiria:1995qz} showed that many of these energy definitions coincide for the Kerr-Schild class of metrics, this restriction on the coordinate choice makes computations for spherically/axially symmetric spacetimes particularly cumbersome (see, e.g.~\cite{Virbhadra:1990vs}). Our definition of the energy as in \eqref{eq:Penrose_charge} is {\em coordinate-independent}.

The result \eqref{eq:KN_charge}, however, is different from   those of \cite{Cohen:1984ue} and \cite{Xulu:2000jf}. In \cite{Cohen:1984ue} the Komar mass was evaluated, while in \cite{Xulu:2000jf} a different energy definition was studied. The Kerr-Newman geometry is not an Einstein space so, as discussed in section~\ref{sec:comparing_charges}, there need be no relation between the charge in \eqref{eq:KN_charge} and the Komar mass. In this case, both the Komar mass $q_{\text{K}}[k]$ and $q_{\text{P}}[K]$ are conserved in time but differ by a factor of 2 in the second term of the final expression in \eqref{eq:KN_charge} (see \cite{Xulu:2000jf} for further discussion).\footnote{The Komar mass is conserved in time for the same reason as the Penrose charge explained above, namely that $f_{(1)}[k]_r=0$ in this example.}

The Reissner-Nordström geometry is recovered by taking the limit $a\to 0$. In this limit the charge becomes 
\begin{equation}\label{eq:RN_Penrose}
    \frac{1}{8\pi} q_P[K]\eval_{a=0} = m-\frac{e^2}{2R_0} \,,
\end{equation}
in agreement with the results of \cite{Tod:1983waa} where Penrose's twistorial energy definition \cite{Penrose:1982wp} was used.
For the Kerr geometry (with $e=0$) the charge reduces to $q_P[K]/8\pi =m$, verifying the fact that for uncharged black holes the energy is contained entirely within the horizon.

As discussed in the introduction, gravitational charges for spacetimes
with asymptotic Killing vectors or Killing tensors are constructed only as integrals at infinity.
One of the advantages of our approach is that expressions like \eqref{eq:KN_charge} give insight into the distribution of a charge throughout spacetime. Above, we have discussed the charge contained in a ball whose radius is larger than the outer horizon of the black hole. Equally, we can consider the interior of the black hole and analyse the distribution of charge there. For example, consider the $a=0$ case (the Reissner-Nordstr\"{o}m solution). We introduce the standard ingoing Eddington-Finkelstein coordinates $(v,r,\theta,\phi)$ via
\begin{equation}\label{eq:EF}
    \dd{v} = \dd{t} + \frac{\dd{r}}{1-\frac{2m}{r} + \frac{e^2}{r^2}} \,,
\end{equation}
in terms of which the metric reads
\begin{equation}
    \dd{s}^2 = - \left( 1 - \frac{2m}{r} + \frac{e^2}{r^2} \right) \dd{v}^2 + 2\dd{v} \dd{r} + r^2 ( \dd{\theta}^2 + \sin^2\theta \dd{\phi}^2) \,.
\end{equation}
There is no coordinate singularity at the horizon in these coordinates and the radial coordinate can be extended to $r\in(0,\infty)$. From \eqref{eq:EF}, the CKY $K$ in \eqref{eq:KN_CCKY} can be written in these coordinates simply as
\begin{equation}\label{eq:KN_CCKY_EF}
    K = r \dd{v} \wedge \dd{r} \,.
\end{equation}
With $a=0$, this CKY 2-form is closed but still satisfies \eqref{eq:KN_KV} where the Killing vector is $k=\partial_v$ in these coordinates. From \eqref{eq:KN_CCKY_EF}, we can calculate the Penrose charge $q_P[K]$ within a ball of any radius $R_0\in(0,\infty)$. The result is the same as in \eqref{eq:RN_Penrose}. That is, the distribution of charge takes the same form both in the interior and exterior of the Reissner-Nordstr\"{o}m black hole. In particular, with $e=0$, we see that the charge of a Schwarzschild black hole contained within a sphere or radius $R_0$ is independent of $R_0$, so that all the charge can be viewed as being localised at the singularity.

\subsection{AdS-Kerr black hole}

We now study an example which showcases the relations between charges discussed in section~\ref{sec:Einstein_spaces}. We will calculate four separate charges associated with the AdS-Kerr solution: $q_{\text{P}}[K]$, $q_{\text{P}}^\Lambda[K]$, $q_{\text{K}}[k]$ and $q_{\text{K}}^\Lambda[K]$. The AdS-Kerr spacetime is a four-dimensional Einstein space, so the relations derived in section~\ref{sec:Einstein_spaces} apply.

The AdS-Kerr black hole \cite{Carter:1968ks}   in Boyer-Lindquist coordinates has the metric
\begin{equation}
\begin{split}\label{eq:AdSKerr_metric}
    \dd s^2 &= -\frac{\Delta_\theta}{\Xi}(1-\lambda r^2) \dd t^2 + \frac{\rho^2}{\Delta_\lambda} \dd r^2 + \frac{\rho^2}{\Delta_\theta} \dd \theta^2 + \frac{r^2+a^2}{\Xi} \sin^2 \theta \dd \phi^2 \\
    &\qquad + \frac{2mr}{\rho^2 \Xi^2} \left( \Delta_\theta \dd t - a \sin^2\theta \dd \phi \right)^2 \,,
\end{split}
\end{equation}
where
\begin{equation}
\begin{split}
    \Delta_\lambda &= (r^2+a^2)(1-\lambda r^2) - 2mr \, ,\\
    \Xi &= 1 + \lambda a^2 \, ,\\
    \Delta_\theta &= 1 + \lambda a^2 \cos^2\theta \,,
\end{split}
\end{equation}
and $\rho$ is given in \eqref{eq:KN_defs}. This metric solves the vacuum Einstein equations \eqref{eq:Einstein} with $\Lambda=3\lambda$.
With $\lambda=0$ it reduces to the Kerr solution, while with $a=0$ it reduces to the AdS-Schwarzschild solution. There are two isometries generated by Killing vectors $\partial_t$ and $\partial_\phi$. 
It also admits a CKY 2-form \cite{OkanGunel:2023yzv}
\begin{equation}
\begin{split}\label{eq:CKY_AdSKerr}
    K &= -\frac{1}{\Xi} \big( r \dd r \wedge \left( \Delta_\theta \dd{t} - a\sin^2\theta \dd\phi \right) \\
    &\qquad + a \sin\theta\cos\theta \dd\theta \wedge \left( a(1-\lambda r^2) \dd{t} - (a^2+r^2) \dd\phi \right) \big) \,,
\end{split}
\end{equation}
which reduces to \eqref{eq:KN_CCKY} for $\lambda=0$. Its divergence gives a particular combination of the Killing vectors:
\begin{equation}\label{eq:AdSKerr_KV}
    \hat{K} = k = \partial_t - a\lambda \partial_\phi \,.
\end{equation}

Let us first calculate the charges $q_{\text{P}}[K]$ in \eqref{eq:non-Penrose_charge_boundary} and $q_{\text{K}}[k]$ in \eqref{eq:Komar_charge_boundary} on a surface $S$ of constant $t$ and $r=R_0$. While the 1-form currents $J_{(1)}[\hat{K}]$ and $f_{(1)}[k]$ do not vanish, we find that their radial component does vanish in this example,  $J_{(1)}[\hat{K}]_r=f_{(1)}[k]_r=0$. As explained in the previous subsection, it follows that the charges $q_{\text{P}}[K]$ and $q_{\text{K}}[k]$ defined on $S$ are constant in time. The components of the 1-form currents along the timelike direction do not vanish, so the charges are not invariant under deforming the surface $S$ in the radial direction.

Firstly, evaluating the Penrose charge gives
\begin{equation}\label{eq:AdSKerr_charge}
    \frac{1}{8\pi} q_{\text{P}}[K] = \frac{m}{\Xi} + \frac{\lambda R_0}{2\Xi} (a^2+R_0^2) \,.
\end{equation}
Taking $\lambda=0$, we recover the charge of the Kerr black hole $q_{\text{P}}[K] =m$ as found in the previous subsection. Taking $a=0$ instead, we find that for the AdS-Schwarzschild geometry $q_{\text{P}}[K] = m+\lambda R_0^3/2$. 

The Komar charge gives
\begin{equation}\label{eq:AdSKerr_Komar}
    \frac{1}{8\pi} Q_{\text{K}}[k] = \frac{m}{2\Xi} - \frac{\lambda R_0}{2\Xi}(a^2+R_0^2) \,,
\end{equation}
which differs from \eqref{eq:AdSKerr_charge}. As emphasised in section~\ref{sec:Einstein_spaces}, on a non-Ricci flat space the Komar and Penrose charges are not in general related.
As expected on general grounds, we find that the charges \eqref{eq:AdSKerr_charge} and \eqref{eq:AdSKerr_Komar} are conserved in time, but depend on the value of $R_0$.

Since the AdS-Kerr geometry is an Einstein space, we can use the generalisations introduced in section~\ref{sec:Einstein_spaces} to build conserved 2-form currents $J_{(2)}^\Lambda[K]$ and $F_{(2)}^\Lambda[k]$ defined in \eqref{eq:JLambda} and \eqref{eq:KomarLambda} respectively. These are co-closed 2-forms so the charges $q^\Lambda_{\text{P}}[K]$ and $q_{\text{K}}^\Lambda[K]$ are invariant under arbitrary deformations of the integration surface $S$.
Evaluating the generalised Penrose charge gives
\begin{equation}\label{eq:AdSKerr_gen_Penrose}
    \frac{1}{8\pi} q_{\text{P}}^\Lambda [K] = \frac{m}{\Xi} \,,
\end{equation}
while the generalised Komar charge is
\begin{equation}\label{eq:AdSKerr_gen_Komar}
    \frac{1}{8\pi} q_{\text{K}}^\Lambda[K] = \frac{m}{2\Xi} \,.
\end{equation}
The charges \eqref{eq:AdSKerr_gen_Penrose} and \eqref{eq:AdSKerr_gen_Komar} are 
independent of the time $t$ and and independent of $R_0$.
Indeed, from the conservation of $J_{(2)}^\Lambda[K]$ and $F_{(2)}^\Lambda[K]$, they are unchanged by arbitrary smooth deformations of the surface on which they are defined. Moreover, we see that they are related by \eqref{eq:Komar=KT}.

Other definitions of the energy of the AdS-Kerr solution appear in the literature, some of which agree with our expressions (e.g.~\cite{Hawking:1998kw} agrees with \eqref{eq:AdSKerr_gen_Penrose}) while others differ. We note that the CKY 2-form \eqref{eq:CKY_AdSKerr} can be multiplied by any non-zero constant to renormalise the charges. In particular, a different choice of normalisation results in the expression for the energy given in \cite{Gibbons:2004ai}. The benefit of that normalisation is that the charge obeys the first law of black hole thermodynamics and also agrees with that of \cite{Ashtekar:1999jx}.

\subsection{Brane charges}

In section~\ref{sec:main_point} we saw that the currents $J_{(p-1)}[\hat{K}]$ are conserved whenever $\hat{K}$ is the divergence of a CKY $p$-form. $\hat{K}$ can be a KY $(p-1)$-form, but this is not necessary. For $p=2$, this implies that $J_{(1)}[\hat{K}]_\mu = G_{\mu\nu} \hat{K}^\nu$ is conserved whenever $\hat{K}$ is the divergence of a CKY 2-form, regardless of whether it is a Killing vector or not. In section~\ref{sec:Kerr-Newman}, an example was studied where $\hat{K}$ is a Killing vector. In this subsection, we discuss brane solutions in 10-dimensional supergravity and a set of CKY $p$-forms which they admit. We will see that there is a family of such forms whose divergences are not KY $(p-1)$-forms but can still be used to calculate a charge which involves the brane tension. For a 0-brane (i.e.~with a 1-dimensional worldvolume), in particular we will find a CKY 2-form whose divergence is not a Killing vector.

\subsubsection{D0 brane}
\label{sec:D0_brane}

As a first example of this type, consider the D0 brane solution of type IIA supergravity in 10 dimensions. With the brane oriented along the timelike $t$ direction, the metric takes the form
\begin{equation}\label{eq:D0_metric}
    \dd{s}^2 = -H(r)^{-1/2} \dd{t}^2 + H(r)^{1/2} (\dd{r}^2 + r^2 \dd\Omega_8 ) \,,
\end{equation}
with $\dd\Omega_8$ the metric on a round unit 8-sphere. The harmonic function $H(r)$ is given by
\begin{equation}\label{eq:H_D0_function}
    H(r) = 1+\frac{T_0}{r^7} \,,
\end{equation}
with $T_0$ a constant.
The isometry along the brane is generated by the Killing vector
\begin{equation}\label{eq:D0_KV}
    k=\partial_t \,.
\end{equation}
It is possible to verify that the 2-form
\begin{equation}
    K = r H(r)^{1/4} \dd{t}\wedge\dd{r} 
\end{equation}
is a closed CKY 2-form; that is, it satisfies \eqref{eq:CKY_eq} with $\tilde{K}=0$. Moreover, its divergence,
\begin{equation}\label{eq:D0_Khat}
    \hat{K} = H(r)^{-3/4} \left( 1 - \frac{3T_0}{4r^7}\right) \partial_t \,,
\end{equation}
is \emph{not} a Killing vector, as can be straightforwardly checked. From the discussion in section~\ref{sec:2form_currents}, the 1-form current $J_{(1)}[\hat{K}]$ is conserved in this case and can be written as the divergence of the 2-form current $J_{(2)}[K]$ as in \eqref{eq:divJ2=J1}. The 1-form current $J_{(1)}[k]$ with $k$ in \eqref{eq:D0_KV} is also conserved since $k$ is a Killing vector, so there are two different conserved 1-form currents.  From \eqref{eq:D0_Khat}, we see that $\hat{K}$ approaches the Killing vector $k$ as $r\to\infty$. This implies that the charges associated with the two 1-form currents agree when evaluated at infinite distance from the brane. 
However, the 1-form current $J_{(1)}[\hat{K}]$  can be written as the divergence of a covariant 2-form and so the associated charge can be written as a surface integral, while this does not appear to be true for the 1-form current $J_{(1)}[k]$.\footnote{We note that $J_{(1)}[k]$ can be written as the divergence of a 2-form asymptotically \cite{Kastor2004}, so the charge contained in the whole spacetime can be calculated with a surface integral at infinite radial distance from the brane, but not at finite distance.}

Let us calculate the charge contained within a surface $S$ at fixed $t$ and $r=R_0$. This surface is not the boundary of any $(d-1)$-dimensional submanifold of the geometry. Furthermore, since $J_{(1)}[\hat{K}]\neq0$, it is not immediately clear that the charge $q_{\text{P}}[K]$ defined on $S$ is conserved in time. However, in this case the projection of $J_{(1)}[\hat{K}]$ onto the radial direction vanishes so the charge is invariant under deformations of the integration surface $S$ in the timelike direction. On the other hand, the component $J_{(1)}[\hat{K}]_t$ does not vanish so the charge is not invariant under radial deformations of $S$. That is, we expect that the charge contained within $S$ is a function of $R_0$ but not of $t$. Indeed, the surface integral evaluates to
\begin{equation}\label{eq:D0_charge_R}
    q_{\text{P}}[K] = \int_{S^8} \star J_{(2)}[K] = \frac{7\pi^4}{15} T_0 H(R_0)^{-1/4} \left( 1+ \frac{T_0}{8R_0^7}\right) \,.
\end{equation}
The charge contained in the full spacetime is then given by the limit $R_0\to\infty$, which is $q_{\text{P}}[K] \to 7\pi^4 T_0/15$. That is, up to a numerical factor, we recover the mass of the brane.

\subsubsection{Other D brane solutions}

Let us now consider other D$p$ branes, with $0\leq p\leq 6$, which are BPS solutions of either type IIA or IIB supergravity in 10 dimensions. The metrics take the form
\begin{equation}
    \dd{s}^2 = H(r)^{-1/2} \eta_{ab} \dd{x^a} \dd{x^b} + H(r)^{1/2} \delta_{\alpha\beta} \dd{x^\alpha} \dd{x^\beta} \,,
\end{equation}
where $x^a$ (with $a=0,\dots,p$) denote the coordinates tangential to the brane, and $x^\alpha$ (with $\alpha=p+1,\dots,9$) denote the transverse coordinates. We denote the transverse distance by $r$, so $r^2 = x^\alpha x_\alpha$.
The harmonic function is
\begin{equation}
    H(r) = 1 + \frac{T_p}{r^{7-p}} \,,
\end{equation}
with $T_p$ a constant. For $p=0$ this reduces to \eqref{eq:H_D0_function}.
The generalisation of the Killing vector \eqref{eq:D0_KV} for the D0 brane is a KY $(p+1)$-form
\begin{equation}\label{eq:Dp_KY}
    f = H(r)^{-(p+2)/4} \dd{x^0} \wedge \dots \wedge \dd{x^p} \,,
\end{equation}
which is tangential to the brane.
For $p=0$ this reduces to the 1-form dual to the Killing vector \eqref{eq:D0_KV}. Since $f$ is a KY form, from the discussion in section~\ref{sec:main_point} there is a conserved $(p+1)$-form current $J_{(p+1)}[f]$ as defined in \eqref{eq:Jn_def}. However, as in the case of $p=0$ studied above, it does not seem that this KY form can be written as the divergence of a CKY $(p+2)$-form. Therefore, the charge associated with the conserved $(p+1)$-form current cannot be written as a surface integral at finite distance from the brane.

Let us instead look for a CKY form, in analogy to the D0 brane. There is a closed CKY $(p+2)$-form 
\begin{equation}
    K = r H(r)^{(1-p)/4} \dd{x^0} \wedge \dots \wedge \dd{x^p} \wedge \dd{r} \,,
\end{equation}
with divergence
\begin{equation}
    \hat{K} = (-1)^{p+1} H(r)^{-(p+5)/4} \left( 1 + \frac{p-3}{4} \frac{T_p}{r^{7-p}} \right) \dd{x^0} \wedge \dots \wedge \dd{x^p} \,.
\end{equation}
One can verify that $\hat{K}$ is \emph{not} a KY $(p+1)$-form, although it is proportional to \eqref{eq:Dp_KY}. Since $\hat{K}$ is the divergence of a CKY form, the current $J_{(p+1)}[\hat{K}]$ is conserved (see section~\ref{sec:higher_rank_currents}). 

We can evaluate the charge $q_{\text{P}}[K]$ defined on a surface $S$ of constant $t$ and $r=R_0$. As in previous examples, the component $J_{(1)}[\hat{K}]_r=0$ so the charge is invariant under deformations of $S$ in the timelike direction. The component $J_{(1)}[\hat{K}]_t\neq0$, so the charge is not invariant under deformations of $S$ in the radial direction. Explicitly evaluating the charge, we find
\begin{equation}
    q_{\text{P}}[K] = \int_{S^{8-p}} \star J_{(p+2)}[K] = c_p T_p H(R_0)^{-(p+1)/4} \left( 1 + \frac{p+1}{8} \frac{T_p}{R_0^{7-p}} \right) \,,
\end{equation}
where the constant prefactor is $c_p = \frac{1}{4}(7-p)^2(8-p) V_{8-p}$, with $V_n = 2\pi^{(n+1)/2}/\Gamma((n+1)/2)$ the volume of the unit $n$-sphere.
As expected, the charge is conserved in time.
The charge of the full geometry is found by taking $R_0\to\infty$, giving $q_{\text{P}}[K]\to c_p T_p$. Up to a numerical factor, this gives the tension of the brane.
In the near-horizon limit, the D-brane metric approaches $\text{AdS}_{p+2} \times S^{8-p}$ and the CKY $K$ is proportional to the volume form of the AdS factor.

\section{Conclusions}
\label{sec:conclusion}

In this work we have studied conserved quantities in gravitational theories defined using CKY tensors.
On spaces admitting a CKY $p$-form, there is a conserved $(p-1)$-form primary current and a $p$-form secondary current whose divergence gives the primary current. Crucially, the secondary current we construct is covariant. Many physically interesting spaces admit CKY tensors, including black hole solutions, Taub-NUT space, and D-brane spacetimes.
Given a submanifold of the correct dimension, integrating the Hodge dual of the current gives a charge. The secondary current can be used to write this charge as a surface integral and the value of the charge is not sensitive to deformations of the surface.

The currents studied here are a generalisation of those in \cite{Kastor2004}, and are equivalent when the CKY form is KY (i.e.~when the CKY is co-closed). One interesting new feature occurs when the divergence of the CKY form is not a KY form. Such CKY forms exist for D-brane solutions, for example. In these cases, the current is conserved due to highly non-trivial integrability conditions following from the CKY equation.

We end by commenting on some future research directions. 
Firstly, we have emphasised throughout that a generic manifold does not admit a CKY form and the results of this paper apply only to a special subset of spaces where these tensors exist.
While there are many interesting examples of such spaces, we would like to be able to discuss the symmetries and conserved quantities of more general spaces. 
For spacetimes that have  asymptotic CKY tensors (i.e.\ ones that have an asymptotic region that approaches a spacetime admitting CKY tensors asymptotically), the construction discussed here gives asymptotically conserved secondary currents that can be integrated over a suitable surface at infinity. Constructions of this type have appeared previously in \cite{Kastor2004, Cebeci2006, Lindstrom:2021qrk, OkanGunel:2023yzv, Jezierski:1994xm, Jezierski:2002mn, Jezierski:2007ym} and in many of these works asymptotic secondary currents have been found in order to calculate charges. These secondary currents are not covariant, but our construction provides covariant asymptotic secondary currents. We will discuss this in a forthcoming publication. Asymptotic symmetries have been linked to soft theorems in QFT as well as memory effects in GR. It would be interesting to explore the implications of our currents in this context.
 
It would also be interesting to generalise the construction of quasi-local charges in e.g.~\cite{Penrose:1982wp, Dougan:1991zz} to the higher-form currents discussed here. In \cite{Penrose:1982wp}, the tensors needed to define charges are not required to exist globally, but they must exist on a particular submanifold on which the charge is to be defined.

\acknowledgments
CMH is supported by the STFC Consolidated Grant ST/X000575/1. MVCH is supported by a President's Scholarship at Imperial College London. UL gratefully acknowledges the hospitality of the Theory Group at Imperial College as well as travel support from ``Längmanska kulturfonden''.

\section*{Appendices}

\appendix

\section{Integrability conditions for CKY tensors}
\label{app:integrability_conditions}

There are many integrability conditions for CKY tensors in the literature. Here we give some that are relevant for our calculations. Their $p=2$ versions can be found in \cite{Lindstrom:2021dpm} and some related results are given in \cite{Lindstrom:2022qjx}.

Let $K$ be a rank-$p$ CKY tensor. The Ricci identity reads
\begin{equation}
    2\nabla_{[\mu}\nabla_{\nu]} K_{\sigma_1\dots\sigma_p} = (-1)^{p-1} p R\indices{_{\mu\nu[\sigma_1}^\lambda} K_{\sigma_2\dots\sigma_p]\lambda}\,.
\end{equation}
Contracting this with $g^{\nu\sigma_1}$ yields
\begin{align}
\begin{split}
    (d-p+1) \nabla_\mu \hat{K}_{\sigma_2\dots\sigma_p} - \nabla^\nu \nabla_\mu K_{\nu\sigma_2\dots\sigma_p} &= (-1)^{p-1}\big( - R\indices{_\mu^\lambda} K_{\sigma_2\dots\sigma_p\lambda} \\
    & \qquad + (-1)^p (p-1) R\indices{_\mu^{\nu\lambda}_{[\sigma_2}} K_{\sigma_3\dots\sigma_p]\nu\lambda} \big) \,.
\end{split}
\end{align}
Upon using the CKY equation \eqref{eq:CKY_eq}, this becomes
\begin{align}\label{eq:reference}
    (d-p)\nabla_\mu \hat{K}_{\sigma_2\dots\sigma_p} - \nabla^\nu \nabla_{[\mu} K_{\nu\sigma_2\dots\sigma_p]} = (-1)^p R\indices{_\mu^\lambda} K_{\sigma_2\dots\sigma_p\lambda} - (p-1)R\indices{_\mu^{\nu\lambda}_{[\sigma_2} } K_{\sigma_3\dots\sigma_p]\nu\lambda} \,.
\end{align}
Now, symmetrising the $\mu$ and $\sigma_2$ indices gives
\begin{equation}
\begin{split}
    (d-p) \nabla_{(\mu} \hat{K}_{\sigma_2)\sigma_3\dots\sigma_p} &= (-1)^p R\indices{^\lambda_{(\mu}} K_{\sigma_2)\sigma_3\dots\sigma_p\lambda} \\
    &\qquad - (-1)^p \frac{p-2}{2} \big( R\indices{_\mu^{\nu\lambda}_{[\sigma_3}} K_{\sigma_4\dots\sigma_p]\sigma_2\nu\lambda} + \mu\leftrightarrow \sigma_2 \big) \,,
\end{split}
\end{equation}
which reduces to eq.~\eqref{eq:KV_integrability} for $p=2$. The right-hand side vanishing is the condition for $\hat{K}$ to be a rank-$(p-1)$ KY tensor. 

Equivalently, one can anti-symmetrise eq.~\eqref{eq:reference} over the $\mu\sigma_2\dots\sigma_p$ indices and subtract the resulting equation from eq.~\eqref{eq:reference}. This gives 
\begin{align}
    (d-p)\left( \nabla_\mu \hat{K}_{\sigma_2\dots\sigma_p} - \nabla_{[\mu} \hat{K}_{\sigma_2\dots\sigma_p]} \right) &= (-1)^p \left( R\indices{_\mu^\lambda} K_{\sigma_2\dots\sigma_p\lambda} - R\indices{_{[\mu}^\lambda} K_{\sigma_2\dots\sigma_p]\lambda} \right) \nonumber \\
    &\quad - (p-1) \left( R\indices{_\mu^{\nu\lambda}_{[\sigma_2}} K_{\sigma_3\dots\sigma_p]\nu\lambda} - R\indices{_{[\mu}^{\nu\lambda}_{\sigma_2}} K_{\sigma_3\dots\sigma_p]\nu\lambda} \right) \nonumber \\
    &= (-1)^p p \left( R\indices{_\mu^{\kappa\lambda}_{[\kappa}} K_{\sigma_2\dots\sigma_p]\lambda} -  R\indices{_{[\mu}^{\kappa\lambda}_{\kappa}} K_{\sigma_2\dots\sigma_p]\lambda} \right) \label{eq:integrability_general}
\end{align}
which, again, reduces to eq.~\eqref{eq:KV_integrability} for $p=2$.

\section{Explicit derivation of conservation for \texorpdfstring{$p=3$}{p=3}}
\label{app:integrability_2nd_rank}

In this short appendix we outline a derivation of the conservation of $J_{(2)}[\hat{K}]$ and of \eqref{eq:divJ2_integrability} in the case where $p=3$; that is, when $\hat{K}$ is a 2-form given by the divergence of a rank-3 CKY tensor. As stressed in section~\ref{sec:main_point}, $\hat{K}$ is not necessarily a KY 2-form. In the main text, this result was given in \eqref{eq:J[Khat]_conserved} with $p=3$, and followed immediately from \eqref{eq:divJn} once the secondary current $J_{(3)}[K]$ had been identified. Here, instead, we derive this result using only integrability conditions of the CKY equation \eqref{eq:CKY_eq}. We will see that this method, while more direct, is considerably more computationally involved.

Taking the divergence of $J_{(2)}[\hat{K}]$, given in \eqref{eq:Jn_def}, we find terms involving a covariant derivative of the Riemann tensor and others involving a covariant derivative of $\hat{K}$. The former can be seen to vanish from a Bianchi identity, while the latter do not immediately vanish unless $\hat{K}$ is a KY 2-form, which is not the general case. One can use the integrability condition \eqref{eq:integrability_general} (with $p=3$), the Bianchi identities and several symmetry properties of the Riemann tensor to write the result as in \eqref{eq:divJ2_integrability}.

Note in particular that no terms containing the Ricci tensor or scalar curvature contribute to the right-hand side of \eqref{eq:divJ2_integrability}. Therefore, let us begin by considering the simpler case where the Ricci tensor vanishes. We will return to the general case below.
In order for $J_{(2)}[\hat{K}]$ to be conserved, the right-hand side of \eqref{eq:divJ2_integrability} must vanish. Our starting point to show this is to note that for $p>2$ the rank-$(p-1)$ current $J_{(p-1)}[\hat{K}]$ can be written in the form
\begin{equation}
\begin{split}\label{eq:J[Khat]_rewrite}
    J_{(p-1)}[\hat{K}]_{\mu_1\dots\mu_{p-1}} &= \frac{p-1}{2} \left( \nabla^\sigma R\indices{^\rho_{[\mu_1}} \right) K_{\mu_2\dots\mu_{p-1}]\rho\sigma} + \frac{1}{4} \left( \nabla^\rho R\right) K_{\rho\mu_1\dots\mu_{p-1}} \\
    &\qquad + \frac{d-p}{2} \Box \hat{K}_{\mu_1\dots\mu_{p-1}} \,,
\end{split}
\end{equation}
where $\Box = \nabla^\mu \nabla_\mu$. This result follows from an integrability condition of the CKY equation \eqref{eq:CKY_eq} found in \cite[eq. (2.9)]{Lindstrom:2022qjx}.
As a check of this identity, consider the simple case where $\hat{K}=0$ for a third rank CKY tensor $K$. The vanishing of $\hat{K}$ implies that the CKY 3-form $K$ is in fact a KY 3-form. For clarity, we denote this KY 3-form by $F$. The identity \eqref{eq:J[Khat]_rewrite} reduces to
\begin{equation}
    0 = \left( \nabla^\sigma R\indices{^\rho_{[\mu_1}} \right) F_{\mu_2]\rho\sigma} + \frac{1}{4} \left( \nabla^\rho R\right) F_{\rho\mu_1\mu_2 }=-\frac{3}{2} \nabla^\rho S_{(3)}[F]_{\rho\mu_1\mu_2}
\end{equation}
in this case, where the conserved current $S_{(3)}[F]$ is given by
\begin{equation}
    S_{(3)}[F]_{\rho\mu_1\mu_2} = R\indices{^\sigma_{[\rho}} F_{\mu_1\mu_2]\sigma} - \frac{1}{3} R F_{\rho\mu_1\mu_2}\,.
\end{equation}
This current was found in a different context in \cite{Lindstrom:2021qrk}.

The rewriting of the current in \eqref{eq:J[Khat]_rewrite} is interesting since it does not depend explicitly on the full Riemann tensor, but only on covariant derivatives of the Ricci tensor and scalar curvature. For example, on Einstein spaces the first two terms on the RHS of \eqref{eq:J[Khat]_rewrite} vanish and the $(p-1)$-form current can be expressed neatly in terms of the Laplacian of components of $\hat{K}$.
In particular, when the spacetime is Ricci flat, \eqref{eq:J[Khat]_rewrite} for $p=3$ reads
\begin{equation}\label{eq:finally}
    (d-3)\Box\hat{K}_{\rho\sigma} + \hat{K}^{\lambda\tau}R_{\rho\sigma\lambda \tau} = 0 \,.
\end{equation}
This result may also be derived from the divergence on the first index of \eqref{eq:reference} making use of the CKY equation \eqref{eq:CKY_eq}.

Next we act on \eqref{eq:finally} with $\nabla^\rho$ and use Ricci flatness and that $\hat{K}$ is co-closed  to find
\begin{equation}
    (d-3)[\nabla^\rho,\Box] \hat{K}_{\rho\sigma} + \nabla^\rho (R_{\rho\sigma\lambda \tau} \hat{K}^{\lambda\tau}) = 0 \,,
\end{equation}
which can be massaged to give
\begin{equation}\label{eq:wegotthereintheend}
    (d-2)\nabla^\rho (R_{\rho\sigma\lambda \tau} \hat{K}^{\lambda\tau}) = 0 \,.
\end{equation}
Finally, we note that in the Ricci flat case \eqref{eq:Jn_expanded} implies that $J_{(2)}[\hat{K}]_{\rho\sigma} = -\frac{1}{2} R_{\rho\sigma\lambda\tau} \hat{K}^{\lambda\tau}$, hence it follows from \eqref{eq:wegotthereintheend} that $\nabla^{\rho} J_{(2)}[\hat{K}]_{\rho\sigma} = 0$. This verifies the conservation of $J_{(2)}[\hat{K}]$ without using the fact that it can be written as the divergence of a 3-form $J_{(3)}[K]$ in the case of a Ricci flat spacetime.

When the Ricci tensor is non-zero, the calculation is more involved. To illustrate we give some of the steps for $p=3$. If we evaluate the current $S_{(3)}[K]$ (where the argument is now a CKY 3-form $K$), we find a useful relation to the $p=3$ version of \eqref{eq:J[Khat]_rewrite}:
\begin{equation}
\begin{split}\label{eq:J[Khat]_rewrite2}
   - J_{(2)}[\hat{K}]_{\mu_1\mu_{2}}+\frac{d-3}{2} \Box \hat{K}_{\mu_1\mu_{2}}&= \frac 3 2 \nabla^\rho S_{(3)}[K]_{\rho\mu_1\mu_{2}}  + \frac{d-3}{2} \Big(2R^{\rho}{}_{[\mu_1}\hat K_{\mu_{2}]\rho}+R\hat{K}_{\mu_1\mu_{2}} \Big)\,
\end{split}
\end{equation}
When taking the covariant divergence of both sides of this relation, the $S_{(3)}[K]$ term vanishes since $S_{(3)}[K]$ is antisymmetric in all indices. The divergence of the $\Box \hat{K}$ term is 
\begin{equation}
    \nabla^{\mu_1} \Box \hat{K}_{\mu_1\mu_2} = - \nabla^{\mu_1} \left( 2R\indices{^\rho_{[\mu_1}} \hat{K}_{\mu_2]\rho} + R_{\mu_1\mu_2\rho\sigma}\hat K^{\rho\sigma} \right)
\end{equation}
When inserted into the divergence of \eqref{eq:J[Khat]_rewrite2}, this results in
\begin{equation}
0=(d-2)\nabla^{\mu_1}J_{(2)}[\hat{K}]_{\mu_1\mu_{2}}
\end{equation}
so that $J_{(2)}[\hat{K}]$ is co-closed as expected. Therefore, we see that the conservation of $J_{(2)}[\hat{K}]$ can be derived using only integrability conditions of the CKY equation \eqref{eq:CKY_eq}, without the knowledge that it can be written as the divergence of a 3-form $J_{(3)}[K]$. This derivation is, however, far more involved.

\bibliographystyle{JHEP}
\bibliography{references}

\end{document}